\documentclass[twocolumn]{aastex631}
\usepackage{amsmath}
\usepackage{multirow,tabularx}

\newcommand\kms{km$\,$s$^{-1}$}
\newcommand\Msol{M$_{\odot}$}

\newcommand{\hi}{H\,{\sc i}}
\newcommand{\hii}{H\,{\sc ii}}

\submitjournal{AAS Journals}

\shorttitle{Two deceptive dwarfs towards Virgo}
\shortauthors{Jones et al.}
\graphicspath{{./}{figures/}}

\begin{document}

\title{AGC~226178 and NGVS~3543: Two deceptive dwarfs towards Virgo}

\correspondingauthor{Michael G. Jones}
\email{jonesmg@arizona.edu}

\author[0000-0002-5434-4904]{Michael G. Jones}
\affiliation{Steward Observatory, University of Arizona, 933 North Cherry Avenue, Rm. N204, Tucson, AZ 85721-0065, USA}

\author[0000-0003-4102-380X]{David J. Sand}
\affiliation{Steward Observatory, University of Arizona, 933 North Cherry Avenue, Rm. N204, Tucson, AZ 85721-0065, USA}

\author[0000-0001-8200-810X]{Michele Bellazzini}
\affil{INAF – Osservatorio di Astrofisica e Scienza dello Spazio di Bologna, Via Gobetti 93/3, 40129 Bologna, Italy}

\author[0000-0002-0956-7949]{Kristine Spekkens}
\affiliation{Department of Physics and Space Science, Royal Military College of Canada P.O. Box 17000, Station Forces Kingston, ON K7K 7B4, Canada}
\affiliation{Department of Physics, Engineering Physics and Astronomy, Queen’s University, Kingston, ON K7L 3N6, Canada}

\author[0000-0002-1821-7019]{John M. Cannon}
\affiliation{Department of Physics \& Astronomy, Macalester College, 1600 Grand Avenue, Saint Paul, MN 55105, USA}

\author[0000-0001-9649-4815]{Bur\c{c}in Mutlu-Pakdil}
\affil{Kavli Institute for Cosmological Physics, University of Chicago, Chicago, IL 60637, USA}
\affil{Department of Astronomy and Astrophysics, University of Chicago, Chicago IL 60637, USA}

\author[0000-0001-8855-3635]{Ananthan Karunakaran}
\affiliation{Instituto de Astrof\'{i}sica de Andaluc\'{i}a (CSIC), Glorieta de la Astronom\'{i}a, 18008 Granada, Spain}
\affiliation{Department of Physics, Engineering Physics and Astronomy, Queen’s University, Kingston, ON K7L 3N6, Canada}

\author[0000-0002-3865-9906]{Giacomo Beccari}
\affil{European Southern Observatory, Karl-Schwarzschild-Strasse 2, D-85748 Garching bei M\"{u}nchen, Germany}

\author[0000-0003-4486-6802]{Laura Magrini}
\affil{INAF - Osservatorio Astrofisico di Arcetri, Largo E. Fermi 5, 50125 Firenze, Italy}

\author[0000-0002-5281-1417]{Giovanni Cresci}
\affil{INAF - Osservatorio Astrofisico di Arcetri, Largo E. Fermi 5, 50125 Firenze, Italy}

\author[0000-0002-9724-8998]{John L. Inoue}
\affiliation{Department of Physics \& Astronomy, Macalester College, 1600 Grand Avenue, Saint Paul, MN 55105, USA}

\author[0000-0002-8598-439X]{Jackson Fuson}
\affiliation{Department of Physics \& Astronomy, Macalester College, 1600 Grand Avenue, Saint Paul, MN 55105, USA}

\author[0000-0002-9798-5111]{Elizabeth A. K. Adams}
\affil{ASTRON, the Netherlands Institute for Radio Astronomy, Oude Hoogeveesedijk 4,7991 PD Dwingeloo, The Netherlands}
\affil{Kapteyn Astronomical Institute, PO Box 800, 9700 AV Groningen, The Netherlands}

\author[0000-0002-6551-4294]{Giuseppina Battaglia}
\affiliation{Instituto de Astrof\'{ı}sica de Canarias, V\'{ı}a L\'{a}ctea s/n 38205 La Laguna, Spain}
\affiliation{Universidad de La Laguna. Avda. Astrofísico Fco. Sánchez , La Laguna, Tenerife E-38205,  Spain}

\author[0000-0001-8354-7279]{Paul Bennet}
\affiliation{Space Telescope Science Institute, 3700 San Martin Drive, Baltimore, MD 21218, USA}

\author[0000-0002-1763-4128]{Denija Crnojevi\'{c}}
\affil{University of Tampa, 401 West Kennedy Boulevard, Tampa, FL 33606, USA}

\author[0000-0003-2352-3202]{Nelson Caldwell}
\affiliation{Center for Astrophysics, Harvard \& Smithsonian, 60 Garden Street, Cambridge, MA 02138, USA}

\author[0000-0001-8867-4234]{Puragra Guhathakurta}
\affiliation{UCO/Lick Observatory, University of California Santa Cruz, 1156 High Street, Santa Cruz, CA 95064, USA}

\author[0000-0001-5334-5166]{Martha P. Haynes}
\affiliation{Cornell Center for Astrophysics and Planetary Science, Space Sciences Building, Cornell University, Ithaca, NY 14853, USA}

\author[0000-0002-0810-5558]{Ricardo R. Mu\~{n}oz}
\affiliation{Departamento de Astronom\'{ı}a, Universidad de Chile, Camino El Observatorio 1515, Las Condes, Santiago}

\author[0000-0003-0248-5470]{Anil Seth}
\affiliation{Department of Physics \& Astronomy, University of Utah, Salt Lake City, UT, 84112, USA}

\author[0000-0002-1468-9668]{Jay Strader}
\affiliation{Center for Data Intensive and Time Domain Astronomy, Department of Physics and Astronomy, Michigan State University, East Lansing, MI 48824, USA}

\author[0000-0001-6443-5570]{Elisa Toloba}
\affiliation{Department of Physics, University of the Pacific, 3601 Pacific Avenue, Stockton, CA 95211, USA}

\author[0000-0002-5177-727X]{Dennis Zaritsky}
\affiliation{Steward Observatory, University of Arizona, 933 North Cherry Avenue, Rm. N204, Tucson, AZ 85721-0065, USA}



\begin{abstract}

The two sources AGC~226178 and NGVS~3543, an extremely faint, clumpy, blue stellar system and a low surface brightness dwarf spheroidal, are adjacent systems in the direction of the Virgo cluster. Both have been studied in detail previously, with it being suggested that they are unrelated normal dwarf galaxies or that NGVS~3543 recently lost its gas through ram pressure stripping, and that AGC~226178 formed from this stripped gas. However, with Hubble Space Telescope Advanced Camera for Surveys imaging we demonstrate that the stellar population of NGVS~3543 is inconsistent with being at the distance of the Virgo cluster, and that it is likely a foreground object at approximately 10~Mpc.
Whereas the stellar population of AGC~226178 is consistent with it being a very young (10-100 Myr) object in the Virgo cluster. Through a re-analysis of the original ALFALFA \hi \ detection we show that AGC~226178 likely formed from gas stripped from the nearby dwarf galaxy VCC~2034, a hypothesis strengthened by the high metallicity measured with MUSE VLT observations. However, it is unclear whether ram pressure or a tidal interaction is responsible for stripping the gas. AGC~226178 is one of at least five similar objects now known towards Virgo. These objects are all young and unlikely to remain visible for over $\sim$500~Myr, suggesting that they are continually produced in the cluster. 

\end{abstract}



\section{Introduction} 
\label{sec:intro}

Blind surveys of neutral hydrogen (\hi) in the local Universe have revealed a plethora of extremely high mass-to-light ratio systems that have few or no stars \citep{Saul+2012,Adams+2013,Cannon+2015}. Optical and inteferometric follow-up observations have demonstrated that many of these systems are likely high velocity clouds associated with the Milky Way or routine gaseous tidal features \citep{Cannon+2015,Bellazzini+2015b,Sand+2015,Adams+2016,Beccari+2016}. However, some may be genuine Local Group dwarfs, similar to Leo T \citep{Irwin+2007} and Leo P \citep{Giovanelli+2013}, while others appear to be unusual objects in the direction of the Virgo cluster \citep{Cannon+2015,Bellazzini+2015,Sand+2015,Adams+2015}. Uncovering what this class of objects are, how they form, and how common they are, is key to understanding the lowest mass regime of the galaxy luminosity function and how stars populate the lowest mass halos that form galaxies. This letter focuses on one such system.

The AGC~226178 and NGVS~3543 system (Figure \ref{fig:HST_CMD}, left) was first identified in the Arecibo Legacy Fast ALFA (Arecibo L-band Feed Array), ALFALFA, survey \citep{Giovanelli+2005,Kent+2008,Haynes+2011} and was followed up with the Jansky Very Large Array (VLA) to map the distribution of \hi \ emission as part of the ALFALFA `Almost Darks' project \citep{Cannon+2015}, a sample of bright extragalactic \hi \ detections that have scarcely visible stellar counterparts. The system lies on the eastern side of the Virgo cluster, approximately 1.4$^\circ$ SE of M~60 and 4$^\circ$ SE of M~87 (about 400~kpc and 1.2~Mpc in projection, respectively). AGC~226178 was originally identified in the ALFALFA survey as an \hi-only detection, with no clear optical counterpart, hence it was included in the `Almost Darks' sample. NGVS~3543, a low surface brightness (LSB) dwarf spheroidal (cyan dash-dot ellipse in Figure \ref{fig:HST_CMD}), was noted at the time as a nearby object, possibly related to the \hi \ detection, but too offset from the \hi \ centroid to be the counterpart \citep{Kent+2008}. The actual stellar counterpart was essentially invisible given the depth of the SDSS \citep[Sloan Digital Sky Survey,][]{York+2000} imaging available at the time.

VLA observations of the \hi \ in the system \citep{Cannon+2015} localized the emission centered on a GALEX \citep[Galaxy Evolution Explorer,][]{Martin+2005} source approximately 1\arcmin \ to the south of NGVS~3543.\footnote{\citet{Cannon+2015} refer to NGVS~3543 as AGC~229166, its original designation. However, we prefer the former to avoid confusion between NGVS~3543 (AGC~229166) and AGC~226178.} This faint, blue, and clumpy object is visible in deeper optical imaging and is highlighted in our Hubble Space Telescope (HST) image by a dashed green ellipse and circles (Figure \ref{fig:HST_CMD}, left panel). Henceforth, we refer to both this blue stellar counterpart and the \hi \ detection as AGC~226178. Based on the SDSS and GALEX imaging available at the time, \citet{Cannon+2015} settled on the interpretation that AGC~226178 is a normal, low-mass, gas-rich galaxy likely falling into the Virgo cluster and that NGVS~3543 is not physically connected to it.

The blue, clumpy appearance of AGC~226178 is reminiscent of another heavily studied object, SECCO~1 \citep{Bellazzini+2015,Sand+2015}, also known as AGC~226067 \citep{Adams+2015}. SECCO~1 was originally identified as an ultra compact high velocity cloud \citep{Adams+2013}, thought to be a candidate for a new low-mass Local Group object. If it were not for the higher recession velocity of AGC~226178 and the close proximity of NGVS~3543, then AGC~226178 might also have been classified this way. The lack of resolved stars in subsequent ground-based observations of SECCO~1 \citep{Bellazzini+2015,Sand+2015,Adams+2015}, and the color-magnitude diagram from HST imaging \citep{Sand+2017}, suggested that it was far beyond the Local Group, most likely in the Virgo cluster. The high metallicity of this object, given its low stellar mass, indicated that it probably formed from gas pre-enriched by a larger galaxy \citep{Beccari+2017}. At present the most likely hypothesis is that SECCO~1 formed from ram pressure or tidally stripped gas either from the group of dwarfs containing VCC~319, 322, and 334 \citep{Bellazzini+2018} or from the M~86 subgroup near the cluster center \citep{Sand+2017}. 
The similarity between SECCO~1 and AGC~226178 (and other blue, star forming clumps in Virgo) suggest that there may be many such systems in or around the Virgo cluster that have been missed previously due to their extreme properties.

Most recently \citet{Junais+2021} revisited AGC~226178 using optical, H$\alpha$ and UV data from NGVS \citep[Next Generation Virgo cluster Survey,][]{Ferrarese+2012}, VESTIGE \citep[Virgo Environmental Survey Tracing Ionised Gas Emission,][]{Boselli+2018} and GUViCS \citep[GALEX Ultraviolet Virgo Cluster Survey,][]{Boselli+2011}. They came to the striking conclusion that the origin of the gas that formed AGC~226178 was NGVS~3543, which they classify as an ultra-diffuse galaxy \citep[UDG; a very LSB galaxy with the stellar mass of a dwarf, but a half-light radius similar to the Milky Way, e.g.][]{vanDokkum+2015}. In this scenario, both objects are assumed to be Virgo cluster members and star formation (SF) in the UDG (NGVS~3543) would have been recently shut off when its gas reservoir was lost through ram pressure stripping by the intra-cluster medium (ICM). This stripped gas then underwent in situ SF, forming the stellar counterpart of AGC~226178. This hypothesis was supported by a faint UV bridge detected between the two objects, and an analysis of the spectral energy distribution of NGVS~3543, which indicated that SF was only recently shut off.

In this letter we present new HST F814W and F606W imaging with ACS (Advanced Camera for Surveys) and MUSE (Multi Unit Spectroscopic Explorer) observations with the VLT (Very Large Telescope), as well as a re-analysis of the original ALFALFA \hi \ data and the VLA \hi \ mapping from \citet{Cannon+2015}. Together these data provide a new perspective on this system and drastically alter the interpretation of its formation mechanism. 

We adopt 16.5~Mpc \citep{Mei+2007} as the distance to the Virgo cluster throughout.

\section{Observations and reduction}

\subsection{HST observations}

AGC~226178 was targeted with ACS in the F606W and F814W filters as part of program 15183 (PI: D.~Sand). The total exposure times were 2120~s and 2180~s for the two filters respectively. A combined F606W and F814W false color image of the system is shown in Figure \ref{fig:HST_CMD} (left). The standard ACS tools in \texttt{DOLPHOT} \citep{Dolphin2000,dolphot} were used to align the individual exposures and perform photometry on all point-like sources.\footnote{\texttt{DOLPHOT} was run with the following parameters: FitSky=1, RAper=4, Force1=0, Align=4, AlignIter=3, ACSuseCTE=1, and standard values of all other parameters.} To select stars from the resulting \texttt{DOLPHOT} catalog we selected all type 1 and 2 (point-like) objects with magnitude uncertainties of less than 0.3 (in both filters) and no photometry flags. We set a crowding limit of 1 mag in the two filters combined, enforced a combined absolute sharpness value of less than $\sqrt{0.075}$, and finally a roundness threshold of less than 1 in both filters. Galactic extinction corrections \citep{Schlafly+2011} were made based on the dust maps of \citet{Schlegel+1998} at the position of each star.  
The typical E(B-V) value was 0.024~mag for both objects.

The completeness limits of each combined field was estimated using artificial star tests in \texttt{DOLPHOT}. We generated $2\times10^5$ artificial stars with F606W magnitudes ranging from 21-30 and F606W-F814W colors in the range $-$1 $<$ (F606W-F814W) $<$2. These were randomly placed over the image and extracted as if real stars. They were then split into color bins and the recovery fraction as a function of F814W magnitude was fit with an error function. The measured limits (from the error function shape parameters) for 50\% and 90\% recovery fractions were then fit with the combination of a horizontal line and a one-sided parabola (e.g. Figure \ref{fig:HST_CMD}, center panels). At $\mathrm{(F606W-F814W)} = 1$ mag  the F814W 90\% and 50\% completeness limits are 26.4 and 26.9~mag, respectively. 

\subsection{VLA observations}

AGC~226178 was observed previously as part of the ALFALFA `almost dark' galaxies sample \citep{Cannon+2015} and was re-reduced for this work. Standard calibration and reduction methods were applied using the Common Astronomy Software Applications package \citep[\texttt{CASA},][]{CASA}. However, this re-reduction used an improved automatic masking of sources during the \texttt{tclean} task. A full description of the data reduction pipeline will be presented in Inoue et al. (in prep.). These data have a channel width of 7.81 kHz ($\sim$1.65~\kms) and a total bandwidth of 8 MHz. The final imaging used Brigg's robust=0.5 weighting to provide a compromise between sensitivity and angular resolution for the detected \hi \ emission. The resulting beam size was 56\arcsec$\times$45\arcsec. During imaging the channels were averaged and re-binned to a velocity resolution of 5 \kms. The resulting rms noise in 5~\kms \ channels is 1.2~mJy/beam. The source mask for AGC~226178 was generated with \texttt{SoFiA} \citep[Source Finding Application,][]{SoFiA,Serra+2015}. \texttt{SoFiA}'s main algorithm adds pixels to a mask based on a signal-to-noise threshold after smoothing to various resolutions both spatially and in velocity. We applied a 3.5$\sigma$ threshold after smoothing with spatial Gaussian kernels with widths of approximately 1 and 2 times the beam diameter and with a box kernel over 20 and 40~\kms \ (4 and 8 channels). A 90\% reliability threshold was applied, which \texttt{SoFiA} estimates based on the apparent flux of negative (presumably spurious) sources. The resulting moment zero map is shown overlaid on a DECaLS \citep[Dark Energy Camera Legacy Survey,][]{Dey+2019} $g$-band image in Figure \ref{fig:BC3_DECaLS_overlay_VLA+ALFALFA} (top-left).

\subsection{MUSE/VLT observations}

Panoramic, integral-field, intermediate-resolution (R=2000-4000) spectroscopy in the wavelength range 4650-9300~\AA \ of a $\simeq1.0\arcmin \times 1.0\arcmin$ field centered on AGC~226178 was acquired with MUSE@VLT \citep{Bacon+2014}, as part of the observing program 0101.B-0376A (P.I: R. Mu\~noz).
Here we provide only an essential description of the process of data reduction and analysis following the procedures of \citet{Beccari+2017}, a detailed description of these procedures will be provided in a dedicated paper (Bellazzini et al. in prep.).
Six $t_\mathrm{exp}=966$~s exposures were acquired, with a dithering scheme based on regular de-rotator offsets, to improve flat-fielding and homogeneity of the image quality across the field. The raw data were wavelength and flux calibrated, and combined into a single stacked data-cube.  Then we searched for individual sources with \texttt{Sextractor} \citep{Bertin+1996} as peaks standing at $\ge 3.0\sigma$ above the background in an H$\alpha$ and white light image\footnote{The width of the point spread function as measured on the white light image on a bright foreground star is $\simeq 0.8\arcsec$~FWHM.}, obtained from the stacked cube by integration in wavelength (over 4650-9300~\AA). Next, we measured the aperture fluxes of the detected sources (with \texttt{Sextractor}, aperture radius $=1.5\arcsec$) in each slice of the cube, with a wavelength step of 1.25~\AA. Ultimately, the fluxes of individual slices were recombined into a 1-D spectrum for each source.
Visual inspection of all the extracted spectra lead us to identify 15 sources (most of which are resolved or marginally resolved) with at least H$\alpha$ in emission in the range of velocity spanned by galaxies in Virgo \citep[$-500 < cz_\odot/\mathrm{km\,s^{-1}} < 3000$, e.g.][]{Mei+2007}. These sources are all coincident with the various components of AGC~226178 shown in Figure~\ref{fig:HST_CMD} (left) and all have a very similar recession velocity, with a mean $\langle cz_{\sun}\rangle=1584$~\kms \ and a standard deviation of $\sigma=4$~\kms. 

\subsection{ALFALFA data}

The ALFALFA \citep{Giovanelli+2005} observations were taken with the 305~m Arecibo telescope in Puerto Rico between 2005 and 2011. The survey adopted a 2-pass drift scan strategy, giving a total effective integration time of 48~s at any given point within the survey footprint. Radio frequency interference was semi-manually identified and flagged before the drift scans were combined into cubes spanning $2.4^\circ \times 2.4^\circ$ on the sky, with 1\arcmin \ pixels and an angular resolution of approximately 4\arcmin. The cube containing AGC~226178 spans approximately $-2000 < cz_\odot\;\mathrm{km\,s^{-1}} < 3000$ with a channel width of $\sim$5~\kms \ and a velocity resolution of 10~\kms \ (the full survey redshift range of $-2000 < cz_\odot\;\mathrm{km\,s^{-1}} < 18000$ was split into 4 overlapping sub-ranges). The cube has an rms noise of 2.4~mJy per channel and a beam size of 3.8\arcmin $\times$ 3.5\arcmin. Further details regarding the ALFALFA survey and its data products can be found in \citet{Haynes+2011,Haynes+2018}. The original detection of the \hi \ emission of AGC~226178 was at a signal-to-noise ratio (SNR) of 10.3 and $cz_\sun = 1581$~\kms.

\section{Results}

\subsection{HST images and CMDs}

\begin{figure*}
    \centering
    \renewcommand{\tabularxcolumn}[1]{m{#1}}
    \setlength\tabcolsep{0pt}
    
    \begin{tabularx}{2\columnwidth}{CCC}
        \multirow{2}*[48ex]{\includegraphics[width=1.1\columnwidth]{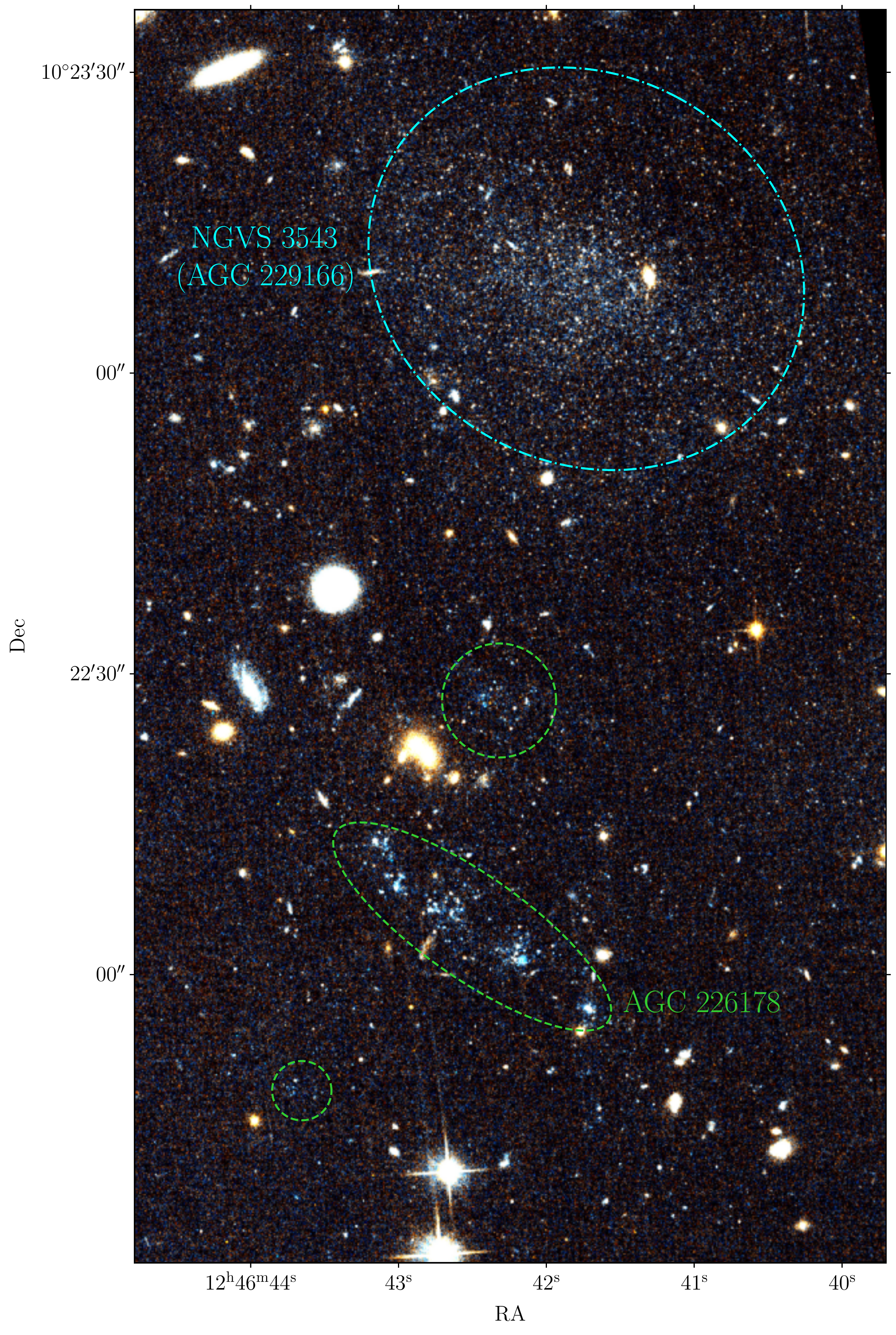}}
            &   \includegraphics[width=0.45\columnwidth]{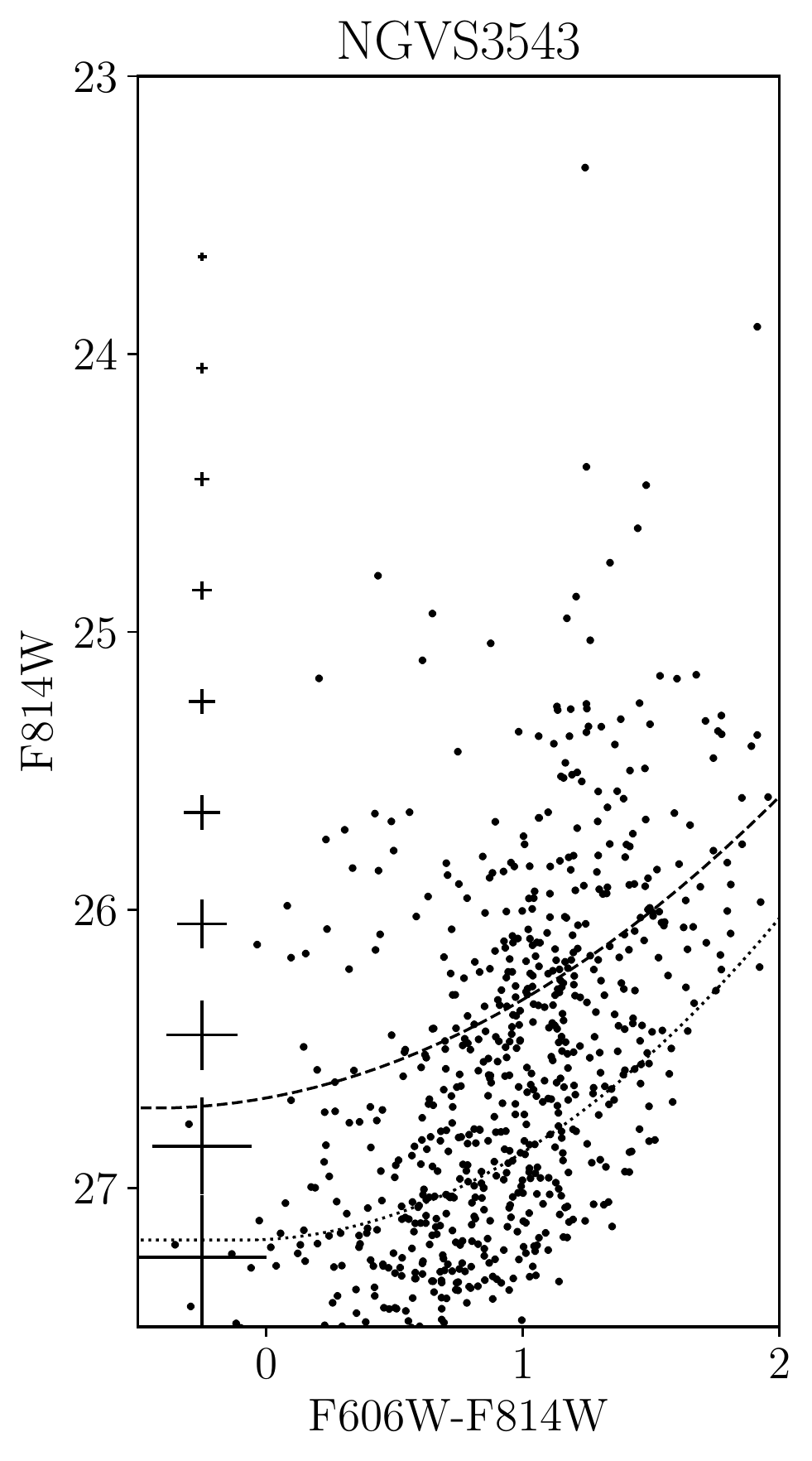}
                &   \includegraphics[width=0.45\columnwidth]{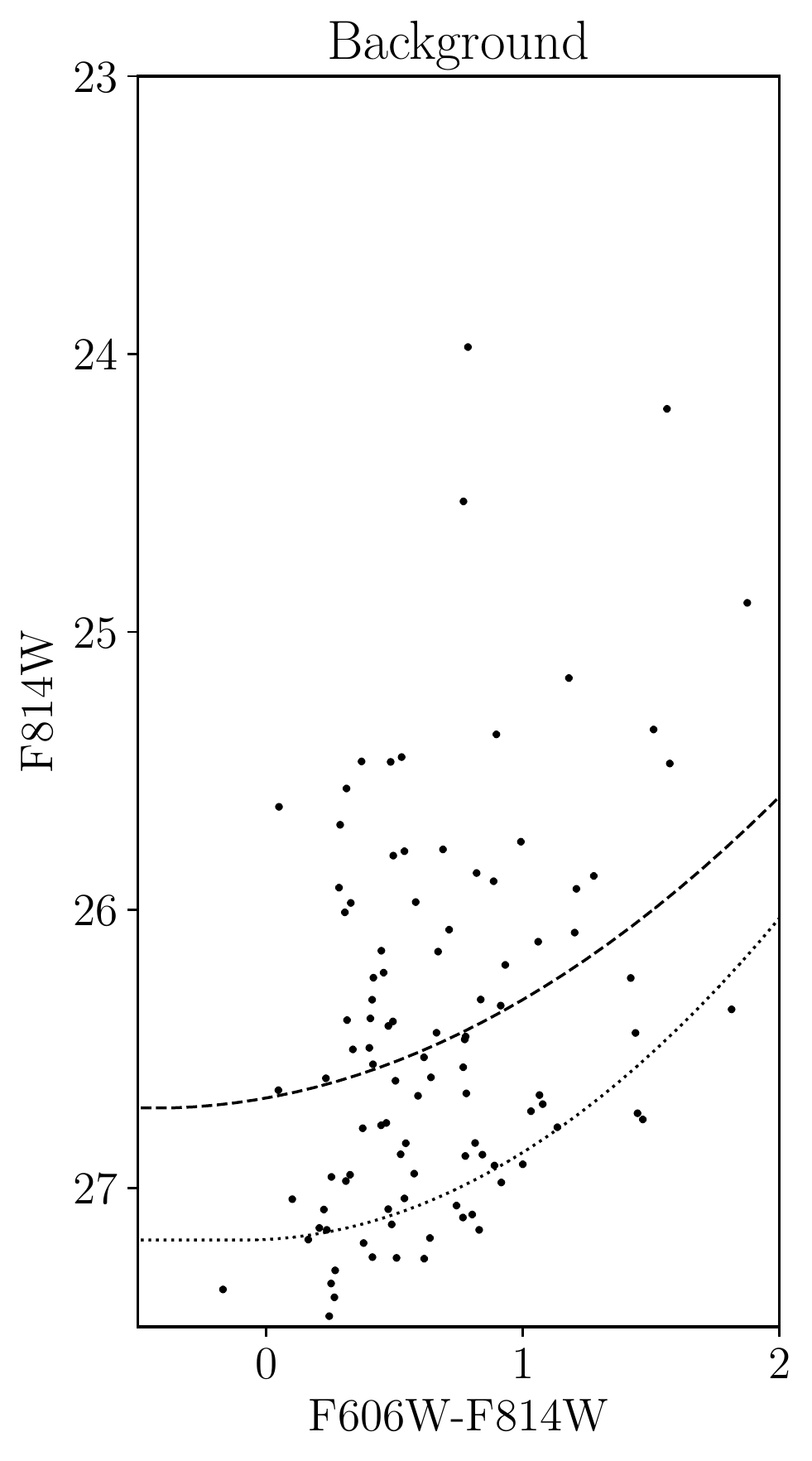}    \\
            &   \includegraphics[width=0.45\columnwidth]{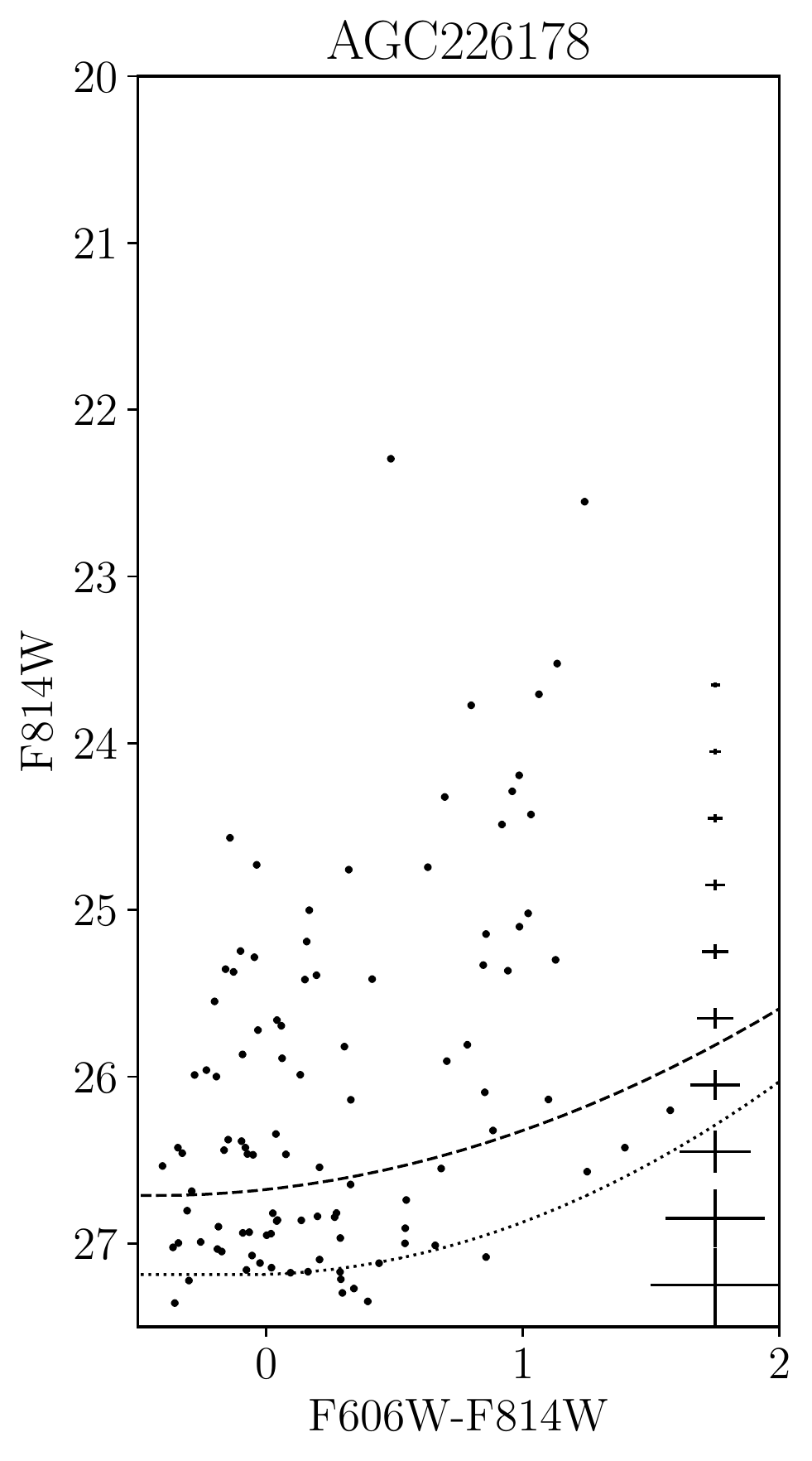}
                &   \includegraphics[width=0.45\columnwidth]{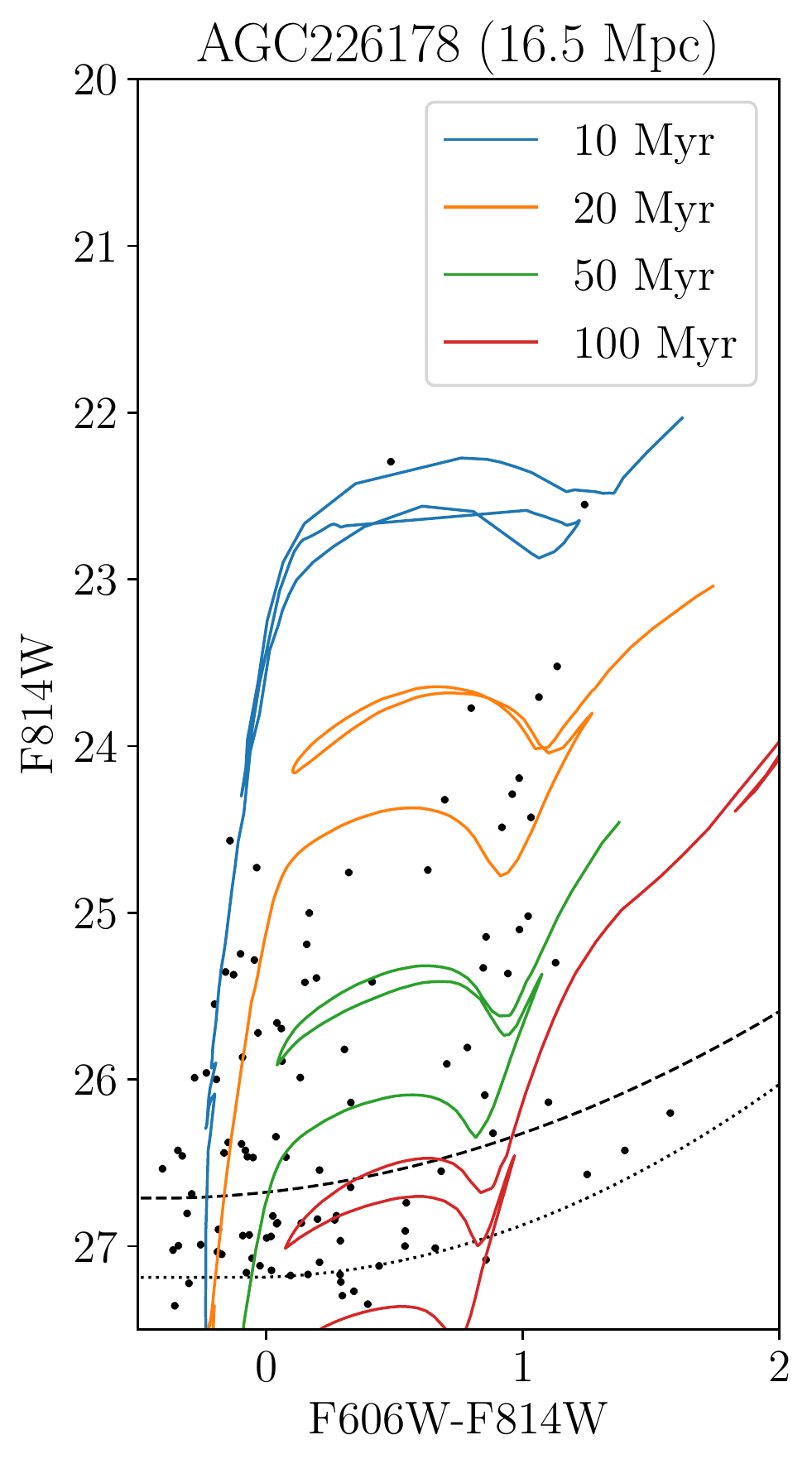}    \\
    \end{tabularx}
    \caption{\textit{Left}: False color F606W+F814W image of the AGC~226178 and NGVS~3543 system. The dash-dot cyan ellipse encircles NGVS~3543 out to the half-light radius \citep[][their table 2]{Junais+2021} and the dashed green ellipse and circles show AGC~226178. \textit{Top-center}: CMD of NGVS~3543 constructed from the stars within two half-light radii (i.e. four times the area of the dash-dot cyan ellipse). The dashed line indicates the 90\% completeness limit and the dotted line the 50\% limit. The errorbars along the edge indicate typical uncertainties in the photometry as a function of F814W magnitude. \textit{Top-right}: CMD of a blank sky region of equal area on the opposite side of the ACS FoV. \textit{Bottom-center}: CMD of AGC~226178 constructed from the stars within the dashed green ellipse and circles. The errorbars along the edge indicate typical uncertainties in the photometry. No background CMD is included, however, the area of AGC~226178 is $\sim$30\% of NGVS~3543 and only $\sim$7 points in the CMD are expected to be background. \textit{Bottom-right}: Repeated CMD of AGC~226178 with stellar population isochrones overlaid for a variety of different ages (assuming the distance to Virgo). Note that the completeness limits in the lower and upper CMDs are the same, but appear different due to the different ranges of F814W magnitudes plotted.}
    \label{fig:HST_CMD}
\end{figure*}

Figure \ref{fig:HST_CMD} shows the HST image of the AGC~226178 and NGVS~3543 system and the CMDs of the two sources. The AGC~226178 CMD contains almost exclusively blue stars while that of NGVS~3543 is dominated by red stars. NGVS~3543 has a fairly smooth elliptical morphology, partially resolved into stars in the HST image, whereas AGC~226178 is decidedly clumpy and appears to be broken into multiple components. The regions (marked by green dashed lines) used to produce the CMD of AGC~226178 were manually constructed after comparison of the HST images and H$\alpha$ detections in MUSE (Section \ref{sec:MUSE_results}). In the case of NGVS~3543 (cyan dash-dot ellipse) the half-light radius and axial ratio from \citet{Junais+2021} were used ($r_\mathrm{e} = 22.38$\arcsec, $i = 30.1^\circ$, $\mathrm{PA} = 61.7^\circ$) to produce the CMD for all stars within 2$r_\mathrm{e}$ (note that the ellipse plotted only extends to 1$r_\mathrm{e}$).

In addition to the AGC~226178 CMD, in the bottom-right panel of Figure \ref{fig:HST_CMD} we overplot \texttt{PARSEC} \citep[PAdova and TRieste Stellar Evolution Code,][]{Bressan+2012} stellar isochrones for a range of ages.\footnote{\url{http://stev.oapd.inaf.it/cgi-bin/cmd}} All isochrones have a metallicity of $\mathrm{[M/H]} = -0.39$ (Section \ref{sec:MUSE_results}) and the assumed distance to Virgo (16.5~Mpc). We note that as isochrones do not depend on the initial mass function (IMF), comparing the CMD to isochrones is robust against an atypical shape of the IMF, which is a possibility for an unusual object such as AGC~226178.

The CMD is complex, but reveals a young stellar population that is similar to the HST CMD of SECCO~1 \citep{Sand+2017,Beccari+2017}. There is a population of faint, blue stars (F814W~$\gtrsim$~24.5, F606W-F814W~$\lesssim$~0) that are likely a combination of young main sequence stars and slightly older blue helium burning stars.  There is also a clear sequence of stars with 23.5~$\lesssim$~F814W~$\lesssim$~26.5 and F606W-F814W~$\gtrsim$~0.6~mag which are likely red helium burning (RHeB) stars. There are no evident RGB stars, which is to be expected if AGC~226178 is at the distance of the Virgo Cluster, as the tip of the red giant branch (TRGB) occurs at $\mathrm{F814W} \approx 27$ at 16.5~Mpc \citep[e.g.][]{Rizzi+2007,Jang+2017}.  The brightness of the RHeB stars can be used to estimate the stellar population age \citep[e.g.][]{McQuinn11}, with the brightest star corresponding to an age of $\sim$10~Myr (at F814W$\approx$22.5~mag), while the faintest corresponds to an age of $\sim$50~Myr.  Older ages are difficult to confirm because the helium burning branch becomes incomplete at fainter magnitudes.  The age spread seen in the CMD is thus similar to that inferred from the GALEX imaging and \hii \ region spectroscopy in the MUSE data (Sections \ref{sec:GALEX_SFRs} \& \ref{sec:MUSE_results}). 

The CMD of NGVS~3543 is discussed in detail in Section \ref{sec:NGVS3543_CMD}. 

\subsection{GALEX Star Formation Rates}
\label{sec:GALEX_SFRs}

AGC~226178 is strongly detected in both NUV and FUV bands (Figure \ref{fig:BC3_DECaLS_overlay_VLA+ALFALFA}, top-right) in GALEX. The total flux in each band was measured from background subtracted GALEX tiles (GI2\_125026\_AGESstrip2\_04 in both NUV and FUV). They were converted to magnitudes following the standard GALEX conversions \citep{Morrissey+2007} and a Galactic extinction correction based on the E(B-V) values of \citet{Schlafly+2011} at the location of each source and $R_\mathrm{NUV} = 8.20$ and $R_\mathrm{FUV} = 8.24$ \citep{Wyder+2007}. Assuming a distance of 16.5~Mpc for all candidates and a bolometric solar absolute magnitude of 4.74, the apparent magnitudes were converted to luminosities and finally to star formation rates (SFRs) following \citet{Iglesias-Paramo+2006}. Uncertainties in the UV fluxes and SFRs were estimated by first masking the brightest 1\% of pixels in the background-subtracted GALEX tiles (to remove bright sources) and then randomly placing $10^4$ circular apertures (of equal area to the aperture in question) across the entire tile in order to estimate the rms noise.

This gives the NUV flux in the combined regions of AGC~226178 as $(1.74 \pm 0.07) \times 10^{-16} \; \mathrm{erg\,s^{-1}\,cm^{-2}\,\AA^{-1}}$, and the FUV flux as $(4.02 \pm 0.09) \times 10^{-16} \; \mathrm{erg\,s^{-1}\,cm^{-2}\,\AA^{-1}}$. These equate to SFR estimates of $\log \frac{\mathrm{SFR_{NUV}}}{\mathrm{M_\odot\,yr^{-1}}} = -3.03\pm0.06$ and $\log \frac{\mathrm{SFR_{FUV}}}{\mathrm{M_\odot\,yr^{-1}}} = -3.18\pm0.06$, respectively. The fact that these two estimates are so similar suggests that the SFR in AGC~226178 has been relatively constant over the past 100~Myr.

NGVS~3543 is weakly detected in GALEX (both NUV and FUV) as pointed out by \citet{Junais+2021}. Making equivalent measurements (to those above) for NGVS~3543 we find only a slightly lower UV flux ($(1.18 \pm 0.19) \times 10^{-16} \; \mathrm{erg\,s^{-1}\,cm^{-2}\,\AA^{-1}}$ in NUV and $(1.47 \pm 0.24) \times 10^{-16} \; \mathrm{erg\,s^{-1}\,cm^{-2}\,\AA^{-1}}$ in FUV), though spread over a much larger area. 
This UV detection (particularly in FUV) would normally imply that SF has not completely ceased, as was suggested by \citet{Junais+2021}. However, the lack of young, blue stars in the CMD (Figure \ref{fig:HST_CMD}, top-center) does not appear to agree with this interpretation. Another possibility is that blue horizontal branch stars, below the completeness limit, are contributing to the UV flux of this object \citep[analogous to][]{Yoon+2004,Goudfrooij2018}. Such stars are known to exist in the old stellar populations of Local Group dwarf spheroidals \citep[e.g.][]{Monaco+2003,Martin+2017} and globular clusters \citep[e.g.][]{Perina+2012,Dalessandro+2012}. However, for NGVS~3543 the color $\mathrm{FUV-NUV} = 0.58 \pm 0.25$, which is slightly bluer than for any of the examples above.

\subsection{\hi \ bridge between AGC~226178 and VCC~2034}

\begin{figure*}
    \centering
    \includegraphics[width=\columnwidth]{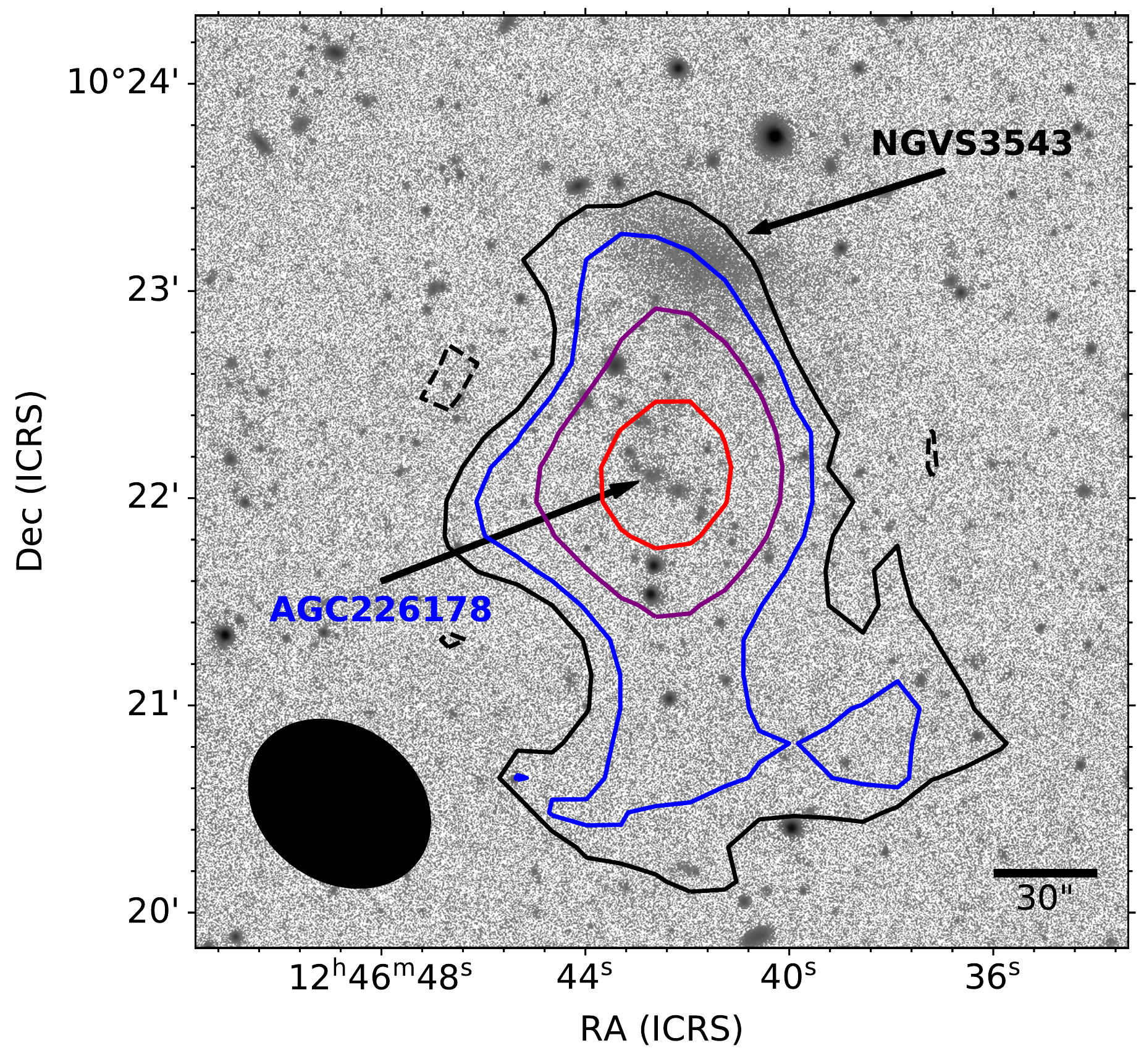}
    \includegraphics[width=\columnwidth]{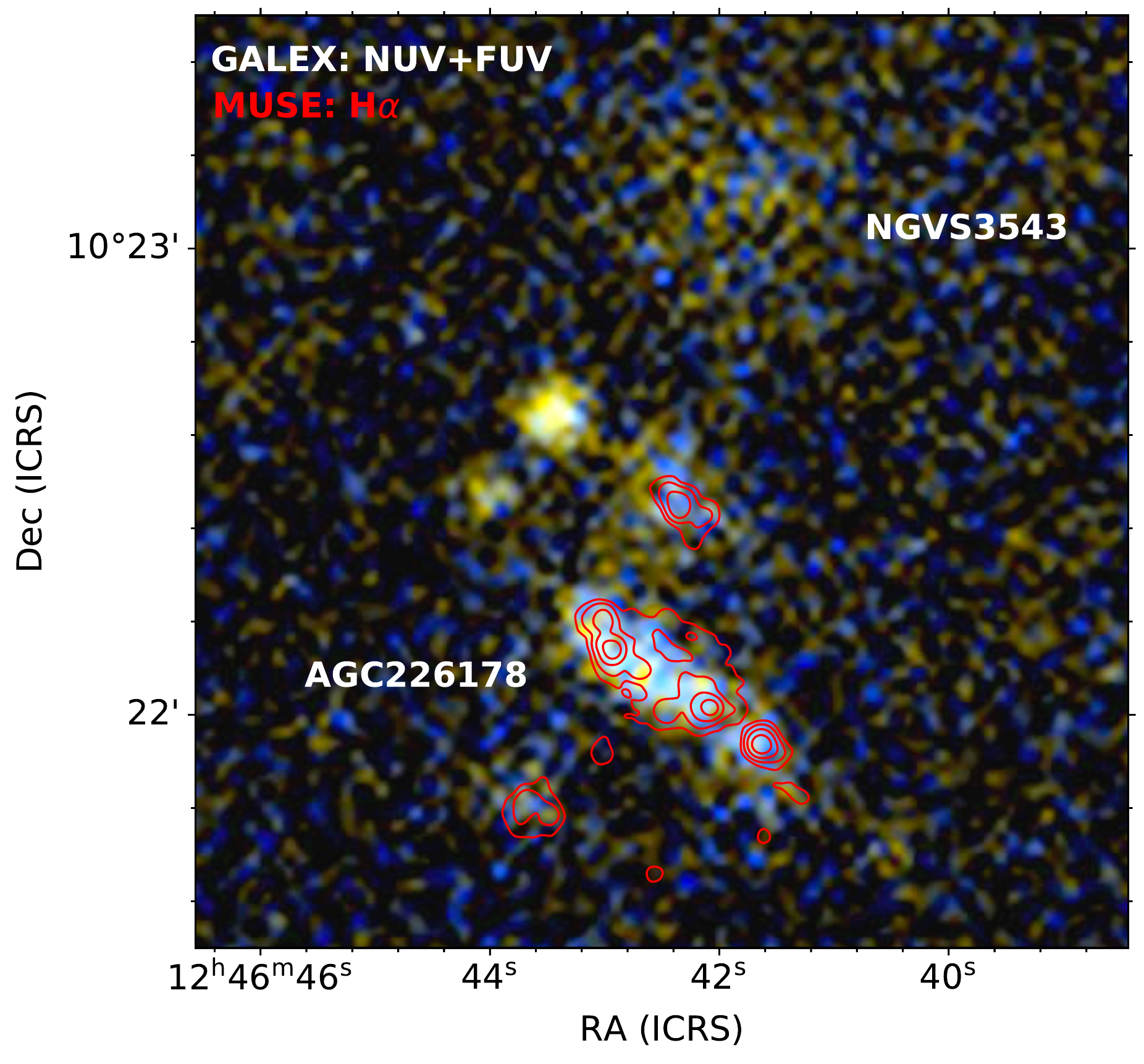}
    \includegraphics[width=\columnwidth]{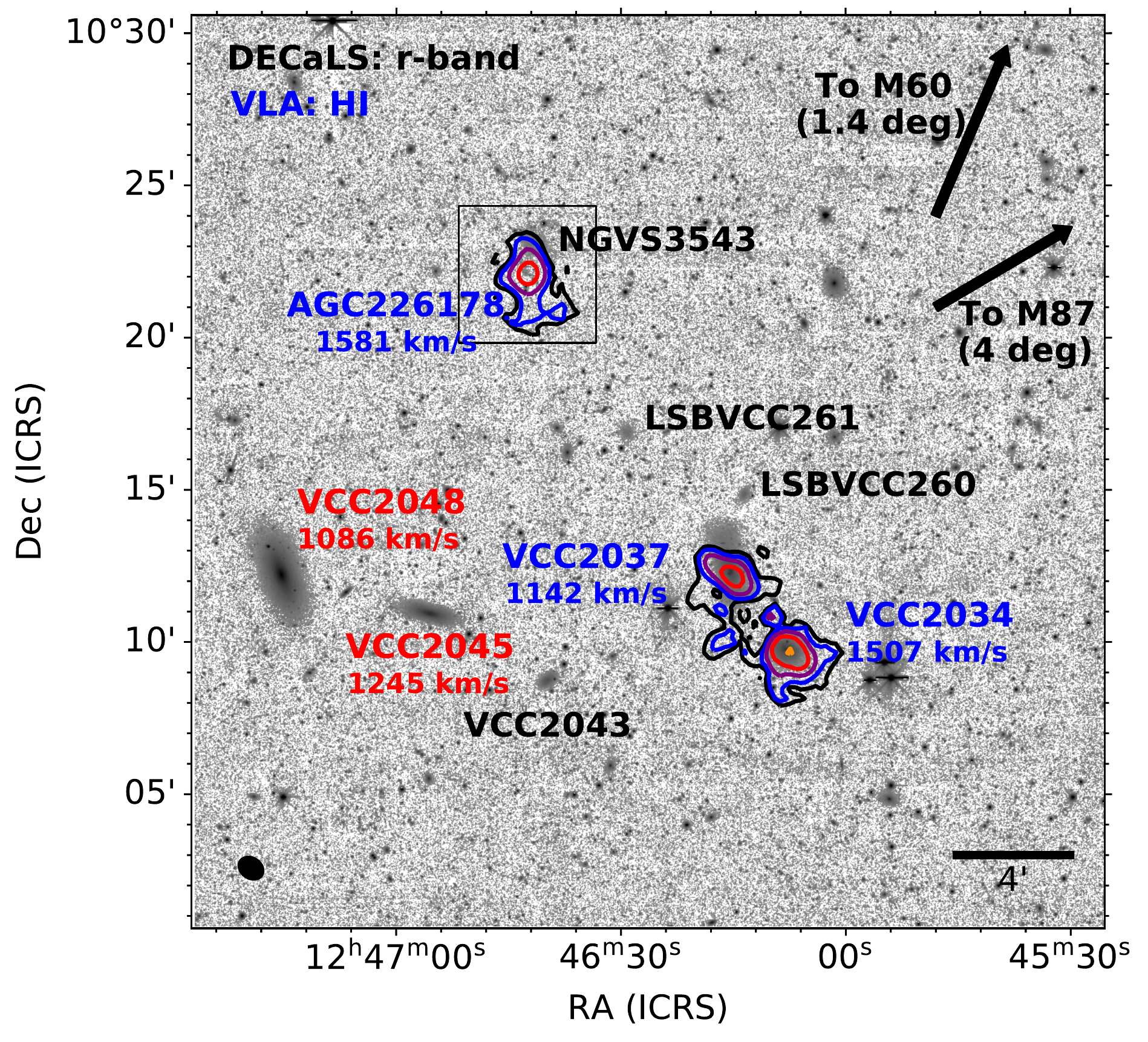}
    \includegraphics[width=\columnwidth]{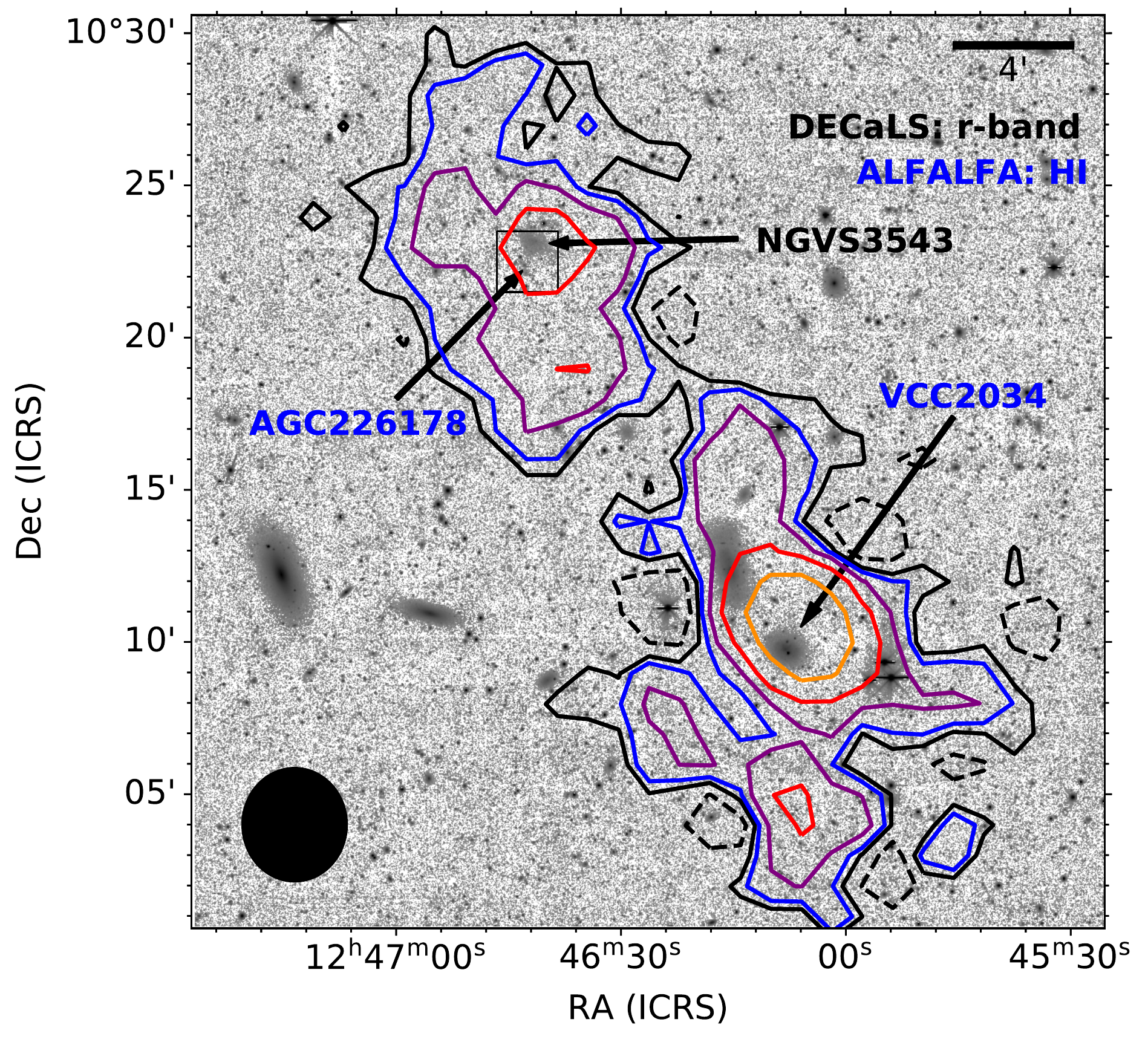}
    \caption{\textit{Top-left}: Contours of integrated \hi \ emission (moment 0 map) from the VLA observations of AGC~226178 overlaid on a DECaLS $g$-band image. The contour levels are -3$\sigma$, 3$\sigma$, 6$\sigma$, 12$\sigma$, 24$\sigma$, and 48$\sigma$. Where 3$\sigma$ corresponds to $1.6 \times 10^{19} \; \mathrm{cm^{-2}}$, or equivalently, $0.13 \; \mathrm{M_\odot\,pc^{-2}}$ (both over 20~\kms). The beam is shown by the black ellipse in the bottom left corner (56\arcsec \ $\times$ 45\arcsec). \textit{Top-right}: GALEX NUV+FUV composite image showing a smaller FoV containing AGC~226178 and NGVS~3543. The red contours show H$\alpha$ emission mapped by MUSE (contour levels are 5, 15, 45, and $135 \times 10^{-20} \mathrm{erg\,s^{-1}\,cm^{-2}}$). AGC~226178 is detected strongly in both GALEX bands and H$\alpha$. NGVS~3543 is $\sim$1\arcmin \ to the north and only weakly detected in GALEX. \textit{Bottom-left}: VLA \hi \ contours overlaid on a DECaLS $r$-band image showing a wide field, including several nearby galaxies that AGC~226178 may have interacted with (contours at the same levels as in the top-left panel). The thin black square outline shows the FoV of the top-left panel. Galaxies with \hi \ detections are labeled in blue, those without \hi \ in red, and those with no redshift measurement in black. \textit{Bottom-right}: ALFALFA \hi \ integrated emission contours overlaid on the same optical image. The thin square outline shows the FoV of the top-right panel. The significance of the contour levels is the same, but in this case 3$\sigma$ corresponds to $2.8 \times 10^{18} \; \mathrm{cm^{-2}}$, or equivalently, $0.02 \; \mathrm{M_\odot\,pc^{-2}}$ (both over 20~\kms). The beam is shown by the black ellipse in the bottom left corner (3.8\arcmin \ $\times$ 3.5\arcmin). In this case the \hi \ emission of VCC~2037 has been excluded to avoid confusion.}
    \label{fig:BC3_DECaLS_overlay_VLA+ALFALFA}
\end{figure*}

\begin{figure*}
    \centering
    \includegraphics[width=0.34\textwidth]{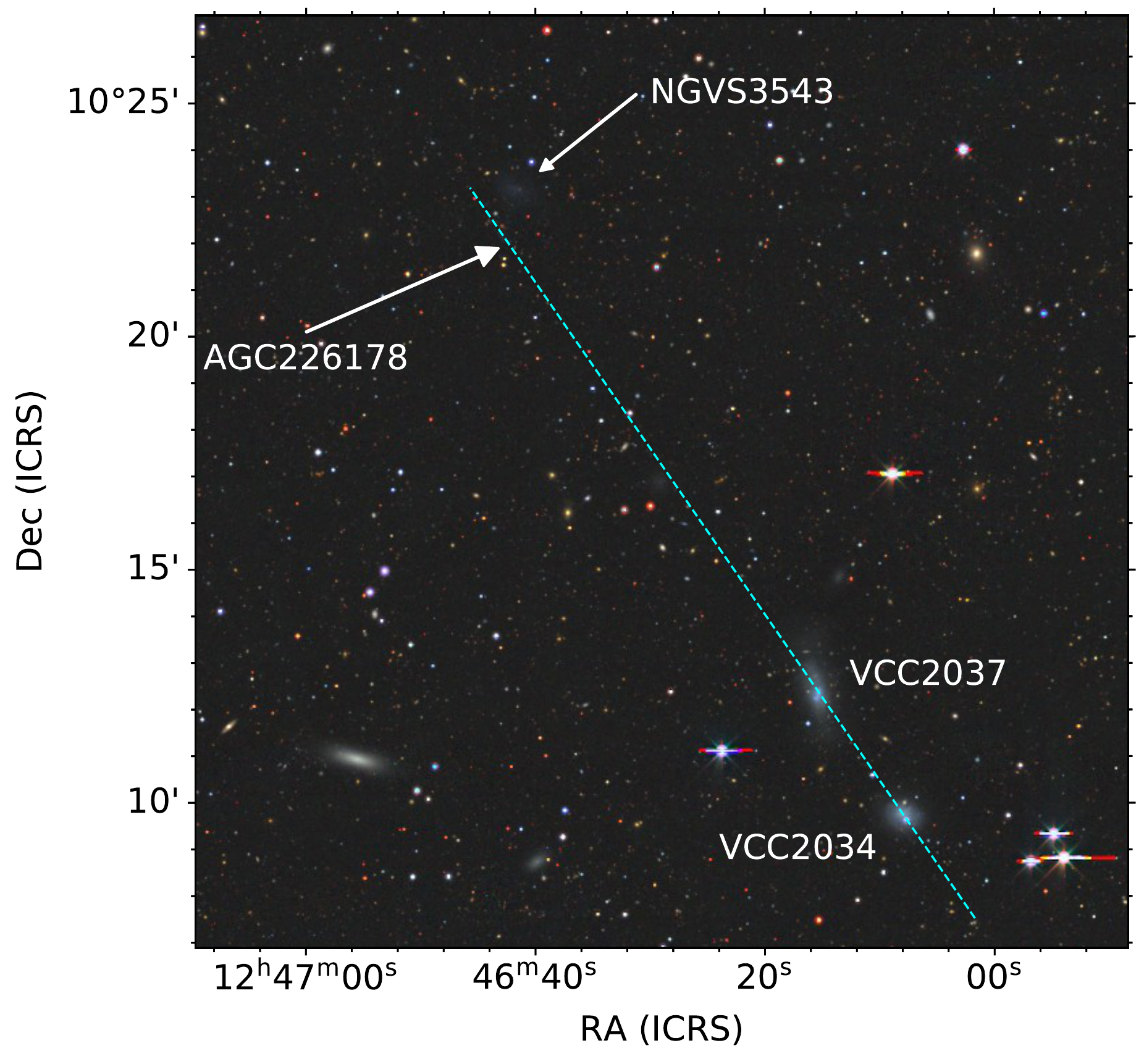}
    \includegraphics[width=0.65\textwidth]{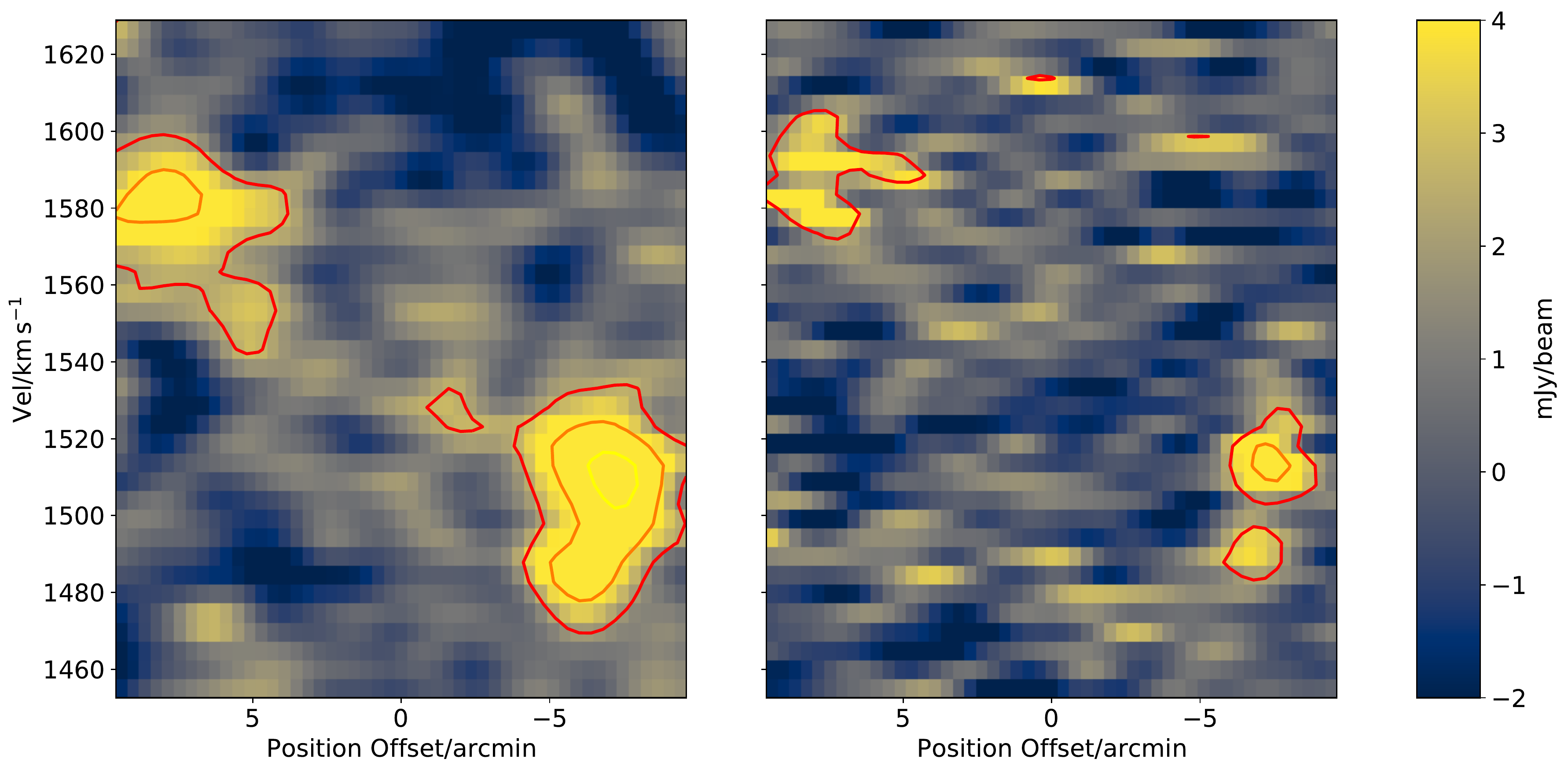}
    \caption{\textit{Left}: DECaLS $grz$ image showing a linear position--velocity slice (cyan dashed line) which intersects with AGC~226178 and VCC~2034. AGC~226178 is not visible on this scale, but its location is marked with an arrow. \textit{Center}: Position--velocity slice from the ALFALFA data cube along the line in the left panel. A region 3.8\arcmin \ wide (centered on the line) was collapsed to form this slice, as this corresponds to the resolution of the ALFALFA data. The contours are drawn at 2.4, 4.8, 7.2 mJy/beam, approximately 1$\sigma$ (red), 2$\sigma$ (orange), and 3$\sigma$ (yellow), respectively. AGC~226178 is the bright clump in the upper left ($\sim1580$~\kms) and VCC~2034 is on the lower right ($\sim1500$~\kms). VCC~2037 does not appear in the slice as it is well below the velocity range plotted. \textit{Right}: The same position--velocity slice, but now from the VLA tapered data cube. A width of 1.8\arcmin \ is used for the slice in this case, again roughly corresponding to the resolution of the data. For clarity the same contour levels are used as in the center panel, however, in this case they correspond to approximately 1.1$\sigma$ and 2.2$\sigma$.}
    \label{fig:pvslice}
\end{figure*}

Figure \ref{fig:BC3_DECaLS_overlay_VLA+ALFALFA} (top-left) shows the VLA \hi \ emission contours of AGC~226178 overlaid on a DECaLS $g$-band image. This map was constructed with \texttt{SoFiA} and extends to lower gas column densities than the equivalent map (based on the same observations) in \citet{Cannon+2015}. This deeper map reveals an extension that appears to point towards the pair of VCC galaxies, VCC~2037 ($cz_\odot = 1142$ \kms) and 2034 ($cz_\odot = 1507$ \kms), to the SW. However, no connection is evident in the VLA data (Figure \ref{fig:BC3_DECaLS_overlay_VLA+ALFALFA}, bottom-left), even though both of these galaxies were detected in the VLA observations of AGC~226178 (about 14\arcmin \ from the pointing center, where the primary beam response is at about 55\%).

The total \hi \ flux of AGC~226178 in the VLA data is 0.3 dex lower than in ALFALFA (0.31~Jy~\kms \ compared to 0.62~Jy~\kms) strongly suggesting that the VLA has missed extended or low column density emission \citep[this was also noted as a possibility by][]{Cannon+2015}. The ALFALFA \hi \ emission is centered at $cz_\odot = 1581$~\kms, which in this direction is consistent with membership of Virgo \citep[e.g.][]{Masters2005}. Assuming a distance of 16.5~Mpc, the larger of these fluxes equates to a total \hi \ mass of $\log M_\mathrm{HI}/\mathrm{M_\odot} = 7.6$. \citet{Junais+2021} estimate the stellar mass of AGC~266178 as $\sim5\times10^{4}$ \Msol, which means the ratio of gas to stellar mass is $\sim$1000 (after multiplying the \hi \ mass by 1.4 to account for Helium). This value is exceptional, even for low-mass, gas-rich galaxies \citep{Huang+2012}, indicating that AGC~226178 is not a normal low-mass galaxy.

Motivated in part by the gas inferred to be missing from the VLA observations we returned to the original ALFALFA cube and constructed a moment 0 map with \texttt{SoFiA}, using smoothing kernels of 1 and 2 times the beam size, 0, 15, and 30~km/s, and a threshold of 3.5$\sigma$ \citep[a reliable threshold for studying extended \hi \ streams, e.g.][]{Taylor+2020}.
The ALFALFA map (Figure \ref{fig:BC3_DECaLS_overlay_VLA+ALFALFA}, bottom-right) shows an apparent \hi \ connection between VCC~2034 and AGC~226178, resolving the question of where the latter's gas likely originated.
We note here that the moment map excludes the velocity range over which there is emission from VCC~2037, so although some contours overlap with it in projection, all the emission is actually associated with VCC~2034 and AGC~226178.

The ALFALFA data has much worse angular resolution than the VLA data, however, it also has significantly better column density sensitivity to extended emission. The majority of the additional emission in the lower-right, versus lower-left panel, of Figure \ref{fig:BC3_DECaLS_overlay_VLA+ALFALFA} is simply below the column density sensitivity of the VLA data and would require a prohibitively long integration time to detect with the VLA (at this resolution). 

To attempt to recover some of this emission in the VLA data, the data were re-imaged after applying a $uv$ taper of $750\lambda$, thereby degrading the resolution of the data to about 2.5\arcmin, but improving the 3$\sigma$ column density sensitivity to $3.2 \times 10^{18}\;\mathrm{cm^{-2}}$ ($0.025 \; \mathrm{M_\odot\,pc^{-2}}$) over 20~\kms, approaching the level of the ALFALFA data. In Figure \ref{fig:pvslice} we show position--velocity slices for both the ALFALFA data and the tapered VLA data, covering the space between AGC~226178 and VCC~2034. 

The \hi \ emission of AGC~226178 is seen as a clump in the upper left of the center and right panels of Figure \ref{fig:pvslice}, and VCC~2034 is the clump in the lower right (of either panel). VCC~2037 is well outside the velocity range plotted. 
The additional emission shown in the ALFALFA moment 0 map (Figure \ref{fig:BC3_DECaLS_overlay_VLA+ALFALFA}, bottom-right) appears to be made up of very low SNR features that extend from both objects towards the other (Figure \ref{fig:pvslice}, center). However, when summed together over several consecutive channels to make the moment 0 map, these combine to form relatively high confidence features. These features may have formed a continuous bridge in the past, or may still do so below the sensitivity of the data. The brightest part of the extension from AGC~226178 is also visible in the VLA position--velocity slice (Figure \ref{fig:pvslice}, right) and there may be a feature extending to lower velocities, but it is at the level of the noise. Although the quoted column density sensitivities of the 2 datasets now only differ by $\sim$25\%, the ALFA beam still covers a 2.5 times larger area. Therefore, if the emission were physically spread over an area larger than the VLA synthesized beam ($\sim$2.5\arcmin), then the effective sensitivity of the VLA data would drop significantly relative to the ALFALFA data. It is also worth noting that the two slices have different widths (Figure \ref{fig:pvslice} caption), again meaning that more extended emission could be missed by the VLA slice. Finally, the quoted column density sensitivity of the VLA data is for the pointing center (approximately the position of AGC~226178). At the location of VCC~2034 the sensitivity is approximately a factor of 2 worse.

As most of the features seen in the ALFALFA data cannot be conclusively corroborated or refuted even with the tapered VLA data, we will primarily consider the ALFALFA data when discussing our interpretation of the system in Section \ref{sec:discuss}.\footnote{We also note that a very similar morphology is seen in WAVES \citep[Widefield Arecibo Virgo Extragalactic Survey;][]{Minchin+2019}, which is approximately 2 times more sensitive that ALFALFA in this region (R. Taylor, private communication).}

\subsection{Velocity and metallicity of AGC~226178 from MUSE}
\label{sec:MUSE_results}

The MUSE observations of AGC~226178 find an abundance of clumps of H$\alpha$ emission. Based on line ratio diagnostics (O\,{\sc iii}/H$\beta$, N\,{\sc ii}/H$\alpha$, and S\,{\sc ii}/H$\alpha$), this emission can all be confidently classified as SF regions \citep[thresholds from][]{Kewley+2021,Kniazev+2008}. The total H$\alpha$ flux measured by MUSE is $3.13 \times 10^{-16} \; \mathrm{erg\,cm^{-1}\,s^{-1}}$. This equates to a total SFR estimate of $\log \frac{\mathrm{SFR_{H\alpha}}}{\mathrm{M_\odot\,yr^{-1}}} = -3.09$ \citep[converted as in equation 2 of][]{Kennicutt1998}, in close agreement with the UV SFR estimates. The mean heliocentric velocity of these \hii \ regions ($cz_\odot = 1584$ \kms) also re-confirms that they are indeed associated with the \hi \ emission. 

Finally, the mean oxygen abundance was estimated by averaging over two different indicators \citep[N2 and O3N2, adopting the calibration by][]{Pettini+2004} for the five sources for which it was possible to measure both indicators \citep[as done by][]{Bellazzini+2018}. These were corrected for extinction based on $\mathrm{H}\alpha/\mathrm{H}\beta$. The average value, $\langle 12 + \log \mathrm{O/H} \rangle = 8.3 \pm 0.1$, supports the finding that the gas likely came from a more massive galaxy that pre-enriched it.\footnote{The quoted uncertainty is the standard deviation between the five sources.} A complete analysis of the abundance measurements will be presented in a separate paper focusing on the MUSE observations (Bellazzini et al. in prep.).

Based on the stellar mass--metallicity relation (MZR) of \citet{Andrews+2013} this metallicity should correspond to a galaxy of $\log M_\ast/\mathrm{M_\odot} = 8.4 \pm 0.4$. VCC~2034 has an $i$-band magnitude of $14.77\pm0.02$ and $g-i = 0.76\pm0.03$ \citep{Kim+2014}. Using the scaling relation of \citet{Taylor+2011} and assuming a distance of 16.5~Mpc, gives $\log M_\ast/\mathrm{M_\odot} = 8.2 \pm 0.1$ for VCC~2034, making it consistent with being the source of AGC~226178's gas.

\begin{table*}
\centering
\caption{Properties of galaxies in the AGC~226178 field}
\begin{tabular}{lcccccccc}
\hline\hline
Object    & RA & Dec & $cz_\odot/\mathrm{km\,s^{-1}}$ & Dist./Mpc      & $g$              & $i$              & $\log M_\ast/\mathrm{M_\odot}$ & $\log M_\mathrm{HI}/\mathrm{M_\odot}$ \\ \hline
AGC226178 & 12:46:42.5 & 10:22:04.8 & 1581                           & (16.5)         & $19.34 \pm 0.05$\tablenotemark{b} & $18.50 \pm 0.06$\tablenotemark{b} & $\sim$4.7\tablenotemark{a}    & $7.60 \pm 0.13$                      \\
NGVS3543  & 12:46:41.7 & 10:23:10.4 &                                & $\sim$10         & $17.50 \pm 0.01$\tablenotemark{c} &     & $6.88 \pm 0.27$\tablenotemark{a}       & $<6.90$                              \\
VCC2034   & 12:46:08.1 & 10:09:45.9 & 1507                           & (16.5)         & $15.53 \pm 0.02$ & $14.77 \pm 0.02$ & $8.2 \pm 0.1$                 & $7.86 \pm 0.08$                      \\
VCC2037   & 12:46:15.3 & 10:12:20.0 & 1142                           & $9.6 \pm 1.0$\tablenotemark{d}  & $15.47 \pm 0.02$ & $14.54 \pm 0.02$ & $8.0 \pm 0.1$                 & $7.18 \pm 0.10$                      \\
VCC2045   & 12:46:55.5 & 10:10:56.7 & 1245                           & (16.5)         & $15.81 \pm 0.02$ & $14.72 \pm 0.02$ & $8.5 \pm 0.1$                 & $<7.60$                              \\
VCC2048   & 12:47:15.3 & 10:12:12.9 & 1086                           & $15.5 \pm 2.0$\tablenotemark{e} & $13.81 \pm 0.02$ & $12.78 \pm 0.02$ & $9.1 \pm 0.1$                 & $<7.55$                              \\ \hline
\end{tabular}
\tablecomments{Columns: 1) Object name. 2) Right ascension (J2000). 3) Declination (J2000). 4) Heliocentric velocity. 5) Distance or assumed distance (if in parentheses). 6) $g$-band magnitude from \citet{Kim+2014} unless indicated otherwise. 7) $i$-band magnitude from \citet{Kim+2014} unless indicated otherwise. 8) Stellar mass estimate based on the $g$ and $i$ magnitudes and the \citet{Taylor+2011} scaling relation (unless stated otherwise). Uncertainties do not include distance uncertainties. 9) \hi \ mass measurements or limits from ALFALFA. Limits are based on the (assumed) distance of each source and the 50\% completeness limit of $\mathrm{S/N} > 6.5$ sources in ALFALFA \citep{Haynes+2011}. A velocity width of 30~\kms \ was assumed for NGVS~3543 and 100 \kms \ for VCC~2045 and 2048.}
\tablenotetext{a}{Estimated by \citet{Junais+2021}, normalized to the relevant distance.}
\tablenotetext{b}{Estimated by summing regions A, B, C, E, and H (which correspond to the clumps we identified as part of AGC~226178) from \citet{Junais+2021}, table 3.}
\tablenotetext{c}{From \citet{Junais+2021}, table 2.}
\tablenotetext{d}{TRGB measurement from \citet{Karachentsev+2014}}
\tablenotetext{d}{Globular cluster luminosity function measurement from \citet{Villegas+2010}. Note there is also a TFR distance estimate (with much larger uncertainty) that places VCC~2048 behind the Virgo cluster \citep{Theureau+2007}.}
\end{table*}

\section{Discussion}
\label{sec:discuss}

In this section we discuss the nature of the AGC~226178 and NGVS~3543 system, focusing in particular on the formation of AGC~226178. The distribution of the \hi \ emission seen with ALFALFA (Figure \ref{fig:BC3_DECaLS_overlay_VLA+ALFALFA}, bottom-right) points to VCC~2034 as the origin of the gas in AGC~226178, and strongly suggests that it formed from stripped gas, making it either a tidal dwarf (TD) or the ram pressure equivalent. The elevated metallicity of the gas, given the stellar mass, also points to the same scenario. 

Without the ALFALFA map, VCC~2034 would seem an unlikely parent galaxy as, based on its ALFALFA flux, it is only $\sim$2 times more massive in \hi \ than AGC~226178. However, as the \hi \ emission appears to connect the two sources and they are at almost exactly the same redshift, it is highly likely that VCC~2034 is the source of the gas. That being the case, there are still two open questions that are essential to the interpretation of this system: 1) what is the distance to the system, and 2) how was the gas stripped from VCC~2034? 

\subsection{Distance to AGC~226178}
\label{sec:dist_to_BC3}

Given its position and velocity, the preferred distance to AGC~226178 is 16.5~Mpc, our assumed distance to objects associated with the Virgo cluster. However, without a direct distance measurement to either AGC~226178 or VCC~2034 (the apparent source of its gas), there remains the possibility that neither are genuinely in Virgo, with the second most likely distance being $\sim$10~Mpc \citep[the distance to VCC~2037, based on a TRGB measurement,][]{Karachentsev+2014}.

Given the morphology of the \hi \ emission it is reasonable to assume that AGC~226178 and VCC~2034 are at the same distance, meaning that a distance estimate to either object would constrain both. We considered estimating a Tully-Fisher relation (TFR) distance to VCC~2034, however, given its low-mass, apparent low inclination, and ongoing interaction, this approach would not be reliable. In the absence of a robust distance estimate, here we note several points in favor of assuming 16.5~Mpc as the distance to this system.
\begin{itemize}
    \item The single strongest argument for $D\approx16.5$~Mpc is that the numerous \hii \ regions detected in the MUSE observations of AGC~226178 mean that it must contain very young stars ($<10$~Myr). When isochrones (assuming $D=16.5$~Mpc) of various ages are overplotted on its CMD (Figure \ref{fig:HST_CMD}, bottom-right), the brightest stars overlap with the 10~Myr isochrone. Moving AGC~226178 to 10~Mpc would shift this isochrone (and all others) approximately 1~mag brighter. Therefore, if AGC~226178 were at 10~Mpc then stars of $\mathrm{F814W}\approx 21.5$ should be present in the CMD. However, there are none. As the completeness is essentially 100\% at this magnitude (and the photometric errors small) it would be difficult to explain the absence of these stars. Whereas a distance of 16.5~Mpc naturally explains the observed population.
    \item VCC~2037 is known to be at $\sim$10~Mpc and it is separated from VCC~2034 by approximately 350~\kms, which would be a large velocity offset if the two were part of the same foreground structure. If that were the case then most likely they would not be gravitationally bound to each other (as both are dwarf galaxies with $M\ast \sim 10^8$~\Msol) and we would be seeing the system at a special time, right as they pass by each other.
    \item If VCC~2034 and VCC~2037 were at the same distance it is likely that they would be interacting and that this interaction would be visible in \hi \ emission, especially given the abundance of loosely bound gas in the vicinity of both galaxies. However, in both the ALFALFA and VLA data cubes there is no clear sign of a bridge or tails extending between the two galaxies.
\end{itemize}

In light of the points above, for the remainder of the discussion we will assume that AGC~226178 is at 16.5~Mpc, in the Virgo cluster. However, at the relevant points we will indicate how our interpretation might change if it were actually at 10~Mpc.

\subsection{Ram pressure or tidal stripping?}

The morphology of the integrated \hi \ emission seen in the ALFALFA data (Figure \ref{fig:BC3_DECaLS_overlay_VLA+ALFALFA}, bottom right) indicates that the gas in AGC~226178 originated in VCC~2034, about 70~kpc to the SW (at the distance of Virgo). There are two main mechanisms by which gas is stripped in this way, tidal stripping and ram pressure stripping. The former can occur in almost any scenario where two or more galaxies (at least one of which contains neutral gas) strongly interact, the latter is only active in a region with a sufficiently dense intergalactic medium, such as a galaxy cluster. If AGC~226178 and VCC~2034 are in the Virgo cluster then both of these mechanisms must be considered.

The one-sided morphology of the \hi \ tail initially suggests ram pressure stripping, as these tails trail in the wake of a galaxy as it falls through the ICM, and are thus one-sided. However, these tails also typically \citep[though not exclusively, e.g.][]{Cramer+2019} point approximately radially away from the cluster center \citep[e.g.][]{Chung+2009}, as galaxies usually fall towards the center. In this case the center of the Virgo cluster is approximately NW, whereas the tail extends to the NE, almost perpendicular.

The tail is approximately 70~kpc long (in projection) and if we assume it is $\sim$200~Myr old (i.e. at least twice as old as the oldest stars we identified), then VCC~2034 would have to have a transverse velocity of $\sim$350~\kms \ for the tail to have been formed by ram pressure stripping. The mean radial velocity of Virgo cluster galaxies is 1138~\kms \ \citep{Mei+2007} and the radial velocity of VCC~2034 is 1507~\kms, which, in this scenario, would make its total velocity relative to the cluster center about $\sim$500~\kms. With this velocity it would still be comfortably bound to the cluster, and the transverse velocity is small enough that it could have acquired this in the past, e.g. if it fell towards the cluster as part of a group. Therefore, ram pressure stripping is still a plausible scenario, despite the direction of the tail.

We also note that a visual inspection of the morphology of VCC~2034 in DECaLS and GALEX imaging does not appear to show a trail of stars accompanying the \hi \ tail. This is the expected behavior of ram pressure stripping as it acts only on the interstellar medium, whereas tidal stripping is gravitational and affects all matter equally. However, this is not conclusive as \hi \ gas is typically much more spatially extended (and therefore more loosely bound) than a galaxy's stellar component, meaning that \hi \ can become heavily disturbed before any disruption of the stars is evident.

The one-sided form of the tail also does not rule out tidal stripping as real examples of tidal stripping of \hi \ seldom follow a simple morphology, and are frequently quite one-sided \citep[e.g.][]{Hibbard+2001}. The main issue with interpreting this \hi \ tail as a tidal feature is the lack of an obvious candidate for the perturber, which we discuss in detail below.

\subsubsection{Which galaxy could have stripped gas from VCC~2034?}

In this sub-section we shall assume that the gas tail formed via tidal stripping and discuss which galaxy might have been responsible.

There are a number of galaxies in the vicinity of the system (Figure \ref{fig:BC3_DECaLS_overlay_VLA+ALFALFA}, bottom-left) that could have stripped gas from VCC~2034. VCC~2048 and VCC~2045 to the SE of AGC~226178 are at $cz_\odot = 1086$~\kms \ and $1245$~\kms, respectively. Although neither of these can be entirely ruled out, if they are as close to VCC~2034 as they appear in projection, and interacting with it, then it would be puzzling why the \hi \ features do not clearly extend towards them. Furthermore, they are separated from AGC~226178 by approximately 500~\kms \ and 300~\kms, respectively, thus unless the encounter occurred at high speed (considered further below), with a large component of the relative velocity along the line-of-sight, then neither of these are likely candidates.

VCC~2037 appears to be adjacent to VCC~2034 as if they are an interacting pair; indeed the distribution of \hi \ in VCC~2037 looks to be disturbed (Figure \ref{fig:BC3_DECaLS_overlay_VLA+ALFALFA}, bottom-left). However, the two galaxies are separated by $\sim$350~\kms \ in velocity and VCC~2037 has both TRGB and TFR distance estimates \citep{Karachentsev+2013,Karachentsev+2014} that place it well in the foreground of the Virgo cluster at about 10~Mpc. Therefore, this pair seems to be a chance projection (although VCC~2034 does not have a redshift-independent distance estimate). We also note that this implies that an additional, separate perturber is needed to explain the morphology of VCC~2037, as ram pressure cannot be the culprit in this case.

The other remaining candidate is NGVS~3543 itself. Although it seems an unlikely candidate for a perturber given its diffuse nature, it is right next to AGC~226178 and (assuming they are at the same distance) has a comparable stellar mass to VCC~2034 \citep{Durbala+2020,Junais+2021}, and therefore should be considered. However, the fact that so many stars are identifiable in its CMD (Figure \ref{fig:HST_CMD}, top-center) suggests that it may be closer than Virgo. At the distance of Virgo the TRGB would be expected to fall at approximately magnitude 27 (F814W). The CMD (Figure \ref{fig:HST_CMD}, top-center) shows many RGB stars brighter than this magnitude and we therefore discount it as a potential perturber. A more detailed discussion of NGVS~3543 follows in the next sub-section.

This leaves us with no strong candidate for the galaxy that perturbed VCC~2034. If both VCC~2034 and AGC~226178 are in Virgo then it is possible that a high speed encounter with another galaxy triggered the \hi \ tail and the formation of AGC~226178, but that galaxy is now sufficiently far away that it is not an obvious candidate. In a cluster relative velocities of galaxies could be many hundreds of \kms. As the radial velocity of AGC~226178 is so similar to VCC~2034 ($c\Delta z < 80$~\kms), it is reasonable to assume that the principal component of the velocity of the encounter would have been transverse to the line-of-sight. The oldest observed stars in AGC~226178 are probably $\sim$100~Myr old and if we assume that the tidal interaction occurred a factor of a few times longer ago, then the perturber might be up to 1 degree away at present.

If we assume that the perturber is at the distance of Virgo and has a stellar mass equal or greater than that of VCC~2034 then we can begin to narrow down the possibilities. We searched the SDSS SpectroPhometric catalog \citep{Strauss+2002,SDSSDR16} for galaxies with redshifts less than 3000~\kms. Using the \citet{Taylor+2011} relation to estimate their stellar masses (based on the $g$ and $i$-band \texttt{cModelMag}s and assuming $D=16.5$~Mpc) we selected only those with $\log M_\ast/\mathrm{M_\odot} > 8.2$, the estimated stellar mass of VCC~2034 (Section \ref{sec:MUSE_results}). This narrows down the possible perturbers to 7 objects: VCC~1948, VCC~2012, VCC~2042, VCC~2045, VCC~2048, VCC~2073, and VCC~2080. As only VCC~1948 and VCC~2048 have redshift-independent distance estimates it is difficult to choose between these objects. But, if we further restrict the velocity range to be within $\sim$100~\kms \ of AGC~226178 or VCC~2034, then VCC~1948 and VCC~2080 would appear to be the best candidates at $cz_\odot = 1610$ and 1532~\kms, respectively. As VCC~2080 is to the NE this would more naturally match the geometry of the system. However, even if this were the perturber, at 49\arcmin \ away, proving a definitive association may not be possible at this stage.

We therefore conclude that while a tidal interaction remains a definite possibility for the mechanism responsible for removing gas from VCC~2034 (and thereby forming AGC~226178), if this system is in the Virgo cluster (as the currently available evidence seems to imply), then there is no strong candidate for the would-be perturber, although there are multiple possible candidates. 

An alternative explanation would be if both VCC~2034 and AGC~226178 were foreground objects at $\sim$10~Mpc. In which case VCC~2037 would presumably be interacting with VCC~2034 and be a strong candidate for the perturber. 
There is another object, similar to AGC~226178, that is known to exist in the foreground of Virgo, namely Coma~P \citep[also known as AGC~229385,][]{Janowiecki+2015,Ball+2018}. Coma~P is the most massive of a system of 3 neighboring objects all detected in \hi \ approximately 8$^{\circ}$ north of M~87. Like AGC~226178, Coma~P's stellar counterpart was first identified through its UV emission in GALEX, and only later confirmed in optical with deep imaging \citep{Janowiecki+2015}. Although Coma~P was originally assumed to be behind the Virgo cluster based on its radial velocity, a recent TRGB distance estimate \citep{Brunker+2019} revealed that Coma~P is in fact much closer ($\sim$5.5~Mpc), placing it in a relatively low density, foreground environment \citep[however, see also][]{Anand+2018}. This case demonstrates that objects similar in appearance to AGC~226178 could be foreground objects. However, as discussed in Section \ref{sec:dist_to_BC3}, a distance of 10~Mpc for AGC~226178 is strongly disfavored (though not ruled out) by the available data. Obtaining a direct distance estimate for VCC~2034 would be the most straightforward means to ruling out this possibility.

\subsection{Distance to, and nature of, NGVS~3543}
\label{sec:NGVS3543_CMD}

\begin{figure}
    \centering
    \includegraphics[width=1\columnwidth]{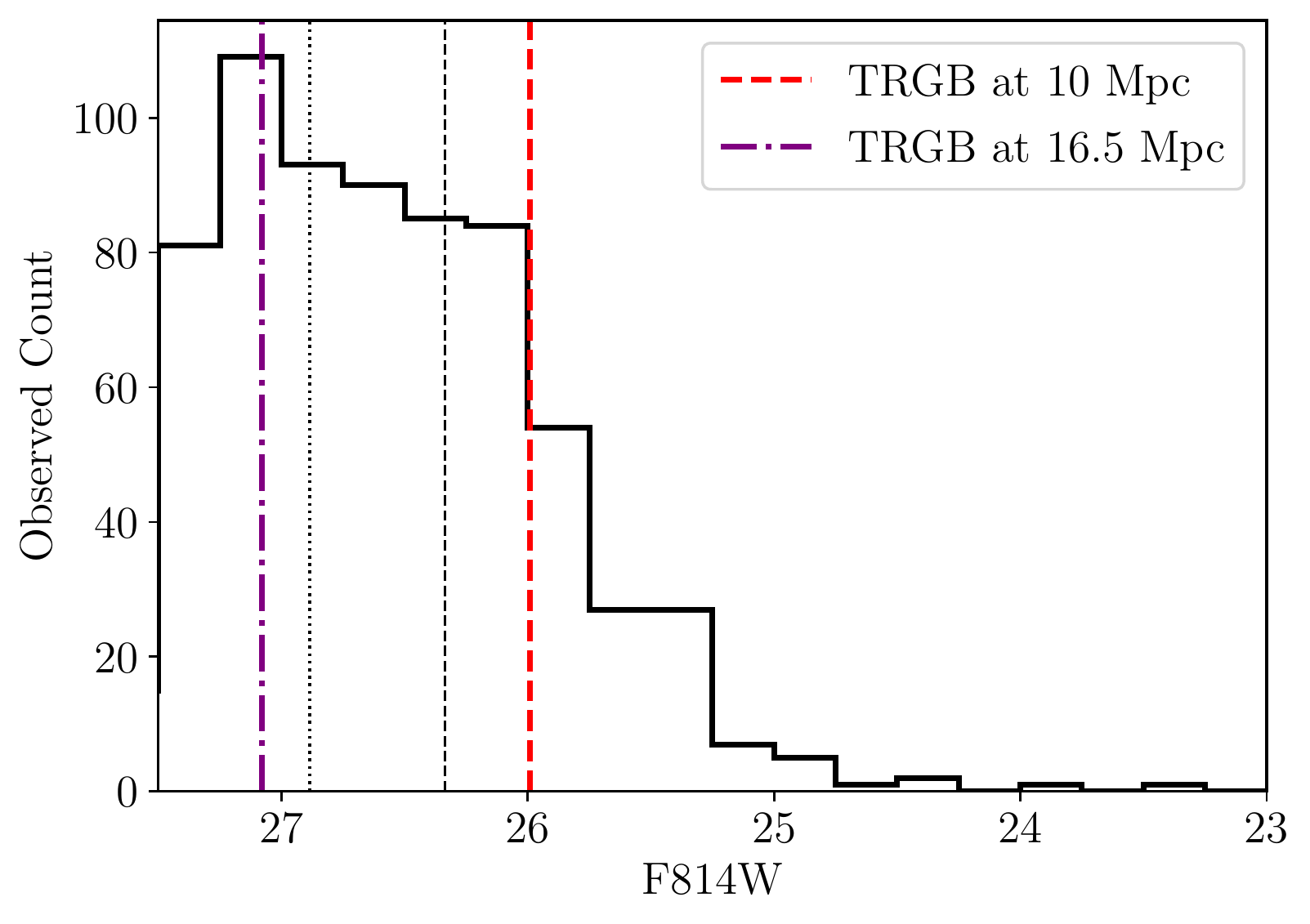}
    \caption{Observed luminosity function of stars in the NGVS~3543 CMD (Figure \ref{fig:HST_CMD}, top center). The thick, vertical, red, dashed line indicates where the TRGB would occur \citep{Jang+2017} for an RGB population with $\mathrm{F606W-F814W} = 1$ and a fiducial distance of 10~Mpc, and the purple dash-dot line corresponds to if NGVS~3543 were at 16.5~Mpc (in Virgo) rather than at 10~Mpc. The vertical dashed line indicates the 90\% completeness limit (at $\mathrm{F606W-F814W} = 1$), and the vertical dotted line the 50\% limit.}
    \label{fig:AGC229166_LF}
\end{figure}

Upon first inspection the close proximity of AGC~226178 and NGVS~3543 might imply that the two objects are physically associated and at approximately the same distance. 
Using the CMD of NGVS~3543 (Figure \ref{fig:HST_CMD}, top center) we attempted to measure the TRGB to obtain an accurate distance estimate. However, these measurements failed to reliably converge, likely owing to the paucity of stars and the proximity of the TRGB to the completeness limit. Instead we turn to the observed luminosity function shown in Figure \ref{fig:AGC229166_LF}. This indicates that the TRGB likely occurs at $\mathrm{F814W}\approx26$ (very close to the 90\% completeness limit), which corresponds to a distance of $\sim$10~Mpc. The stars brighter than this in the CMD likely belong to an AGB population \citep[analogous to the case in][]{Sand+2014}. We also find that the distribution of these candidate AGB stars is considerably more compact than that of the candidate RGB stars, suggesting that they may be a younger population.
The CMD of NGVS~3543 can also be compared to another dwarf spheroidal, known to be in the Virgo cluster, Dw~J122147+132853 \citep[][their figure 6]{Bellazzini+2018}. The majority of the detected stars in this dwarf are fainter than $\mathrm{F814W} = 27$, and there are almost no stars brighter than $\mathrm{F814W} = 26$. This comparison reaffirms that NGVS~3543 is significantly nearer than the Virgo cluster.

Our distance estimate of $\sim$10~Mpc contradicts \citet{Junais+2021}, who assumed that AGC~226178 and NGVS~3543 were at the same distance (in the Virgo cluster) and physically interacting due to a faint bridge of UV emission between them (their figure 1). However, with the HST imaging (and MUSE observations) it is clear that there is a clump of blue stars projected between the main body of AGC~226178 and NGVS~3543 (dashed green circle near the center of the left panel of Figure \ref{fig:HST_CMD}), which likely explains this apparent bridge of UV emission. Given that AGC~226178 is highly clumpy and irregular, this configuration could easily have occurred by chance.

\citet{Junais+2021} also estimated the mass of NGVS~3543 (without a ram-pressure stripping event) as $\log M_\ast/\mathrm{M_\odot} = 7.31 \pm 0.27$ (if it were in Virgo), which if it follows the MZR of \citet{Kirby+2013} would mean it has $\langle [\mathrm{Fe/H}] \rangle = -1.3 \pm 0.2$.\footnote{Here we do not use the relation from \citet{Andrews+2013} as NGVS~3543 is below the mass range covered by their sample.} Assuming a solar ratio of Fe/O \citep{Asplund+2009} this equates to $12 + \log \mathrm{O/H} \approx 7.4$. This value is almost an order of magnitude lower than that of AGC~226178 (Section \ref{sec:MUSE_results}), reinforcing that it is unlikely that the gas in AGC~226178 originated in NGVS~3543 \citep[as hypothesized by][]{Junais+2021}. Moving NGVS~3543 to 10 Mpc would reduce the stellar mass estimate of \citet{Junais+2021} to $\log M_\ast/\mathrm{M_\odot} = 6.88 \pm 0.27$, its absolute magnitude to $M_g = -12.5$, and its metallicity estimate to $\langle [\mathrm{Fe/H}] \rangle = -1.4 \pm 0.2$, further exacerbating the discrepancy. 

At a distance of $\sim$10~Mpc it is likely that NGVS~3543 is associated with VCC~2037 approximately 12\arcmin \ (36~kpc) to the SW, which has a prior TRGB distance estimate of 9.6~Mpc \citep{Karachentsev+2014}. Although at 10 Mpc NGVS~3543 would not quite be classified as a UDG (as its half-light radius would be $<1.3$~kpc), it is similar in appearance. Tidal stripping is a promising mechanism for forming such galaxies \citep[e.g.][]{Bennet+2018,Carleton+2019,Jiang+2019,Liao+2019,Jones+2021} and thus its association with VCC~2037 might explain its appearance. Furthermore, it is possible that an interaction with VCC~2037 led to the recent quenching of this object that \citet{Junais+2021} proposed.
Alternatively, NGVS~3543 may simply be a field dwarf at a slightly different distance to VCC~2037. However, if this were the case it would be an exceptionally odd system, as it has a diffuse morphology, UV emission, but no detectable \hi, and its CMD is dominated by red stars.

\subsection{Fate of AGC~226178}

The position of the stars in the CMD of AGC~226178 (Figure \ref{fig:HST_CMD}, bottom-right) indicates that there is likely some spread in their ages in the approximate range 10-100~Myr. This would also be consistent with the close agreement in the SFR estimates from NUV and FUV, suggesting that the SFR has been somewhat constant on the order of 100~Myr. Furthermore, a sustained SFR of $\sim$10$^{-3}\;\mathrm{M_\odot\,yr^{-1}}$ (Section \ref{sec:GALEX_SFRs}) over 100~Myr, would produce a total stellar mass of $\sim$10$^{5}$~\Msol, similar to the stellar mass estimate of \citet{Junais+2021}, $5\times10^4$~\Msol. At its present SFR, AGC~226178 could go on producing stars for several Hubble times without running out of gas. However, it is unlikely that it will be able to retain its gas on a long ($>1$~Gyr) timescale. 

As it appears to have formed from stripped material, AGC~226178 is unlikely to contain a significant quantity of dark matter, and as its total baryonic mass is less than $10^8$~\Msol \ it does not meet the threshold typically assumed for a TD to be long-lived \citep{Bournaud+2006}. This means that AGC~226178 is unlikely to be self-gravitating, and even if it is, then it is unlikely to remain so on a Gyr timescale. Therefore, it may not be appropriate to refer to it as a galaxy \citep{Willman+2012}, and we have tried to avoid doing so. However, unlike in field and group environments, where TDs are typically studied, it has been suggested that the hot ICM may actually assist low-mass, gas-rich objects in remaining bound longer than would be expected from their internal gravity alone, as this can add an additional external compression to the gas \citep{Burkhart+2016}. 

The fate of the \hi \ component of a very similar object (SECCO~1) was investigated in detail by \citet{Bellazzini+2018} and \citet{Calura+2020}. These works performed a series of high resolution hydrodynamical simulations of a $10^7$~\Msol \ gas cloud moving at up to 400~\kms \ through a hot medium, finding that a significant fraction of its initial \hi \ content could persist for on the order of 1~Gyr. However, instabilities in the gas cloud build up as it moves through the hot ICM and on longer timescales it will be broken up into smaller clouds and evaporate. As \hi \ makes up the vast majority of the baryonic mass of AGC~226178, and provides the medium for external pressure equilibrium, once the \hi \ is lost it will almost certainly be unbound.

Given its low stellar mass, AGC~226178 is only visible at all in the optical because of its young, bright, blue stars. Once SF ceases it will rapidly fade. After 100~Myr only a handful of stars would be above the 90\% completeness limit of our HST observations, and after 500~Myr almost no individual stars would be detectable at all. Assuming its total stellar mass has not dramatically increased in the intervening time (such that its integrated light would be detectable), then it will become, for all intents and purposes, invisible \citep[c.f.][]{Roman+2021}, and its aging stars will become part of the intra-cluster light of Virgo.

\subsection{A larger population of blue stellar systems in Virgo}

As alluded to by \citet{Sand+2017}, SECCO~1 and AGC~226178 are not the only objects of their kind. Both were originally identified by their \hi \ emission in the ALFALFA survey and were noteworthy due to their near invisible (in SDSS images) stellar counterparts. As part of this work, AGC~226178 was re-identified (independent of it original \hi \ detection) during a visual search for faint, blue, and UV-emitting objects in the Virgo cluster using $\sim$100~deg$^2$ of publicly available NGVS images and GALEX tiles. A full analysis of this search will be presented in a subsequent paper (Jones et al. in prep.), but at least 3 additional objects were identified with very similar optical and UV properties to AGC~226178. Assuming the other objects are also in Virgo and extremely young (as AGC~226178 appears to be) then it is reasonable to assume that such objects are being continually created in the Virgo cluster. At present these $\sim$5 objects are known and, based on the discussion above, each one may only be visible for $\sim$500~Myr. Therefore, these objects must be being produced in Virgo at a rate of $\sim$1 per 100~Myr. However, it remains to be seen whether these form a uniform population with related formation mechanisms, or if they are really a mixture of many different types of object that happen to have similar appearances in optical and UV.

\section{Conclusions}

The AGC~226178 and NGVS~3543 system is a disparate (false) pair of dwarf galaxies in the direction of the Virgo cluster, where foreground and background objects are frequently projected next to each other. With a re-analysis of ALFALFA and VLA data (Figures \ref{fig:BC3_DECaLS_overlay_VLA+ALFALFA} \& \ref{fig:pvslice}) we have demonstrated that AGC~226178 likely formed from gas stripped from VCC~2034, and that it is neither a normal gas-rich dwarf \citep{Cannon+2015} nor a stellar system formed through the ram pressure stripping of a UDG \citep{Junais+2021}. The hypothesis is also supported by its high metallicity, indicating its gas originated in a galaxy with a much higher stellar mass. Its apparent neighbor, NGVS~3543, appears to be a foreground object at approximately 10~Mpc, based on its CMD (Figures \ref{fig:HST_CMD} \& \ref{fig:AGC229166_LF}). At this distance it would be too small to meet the classification criteria of a UDG. 

The magnitude of the youngest, brightest stars in AGC~226178 is consistent with this object being in the Virgo cluster, though a direct distance measurement to either AGC~226178 or VCC~2034 would solidify this assessment. As both NGVS~3543 and VCC~2037 are foreground objects, there is no obvious candidate for the perturbing galaxy that stripped gas from VCC~2034 to form AGC~226178. This leaves two possible scenarios: 1) a high speed, close encounter, with the perturber now as much as a degree away, or 2) ram pressure stripping with VCC~2034 falling into Virgo with a large transverse velocity ($\sim$350~\kms).

AGC~226178 (and SECCO~1) are part of a larger sample of similar objects \citep[e.g.][]{Sand+2017} in Virgo, or observed towards Virgo, that are similar in optical appearance and which we will present in a upcoming paper. These objects may really be a mixed collection of objects with differing formation mechanisms or may all have formed recently from stripped gas, like AGC~226178. Over the next several hundred Myr AGC~226178 (and other objects like it) will lose its neutral gas, stop forming stars and fade from view, eventually becoming essentially invisible. Thus, for these objects to be visible at any given time they must be being continually produced in the cluster.

Our findings highlight the complexity of studying systems in the direction of Virgo.
In this field there appears to be three close pairs (AGC~226178/NGVS~3543, VCC~2034/2037, and VCC~2045/2048), yet neither of the first two are genuine pairs. Instead, AGC226178/VCC2034 and NGVS3543/VCC2037 both form wide pairs ($\sim$15\arcmin \ separation), but their members are projected into the two different, false, close pairs indicated above.
In addition, both VCC~2034 and VCC~2037 have disturbed morphologies, but do not appear to be interacting with each other, while the actual perturber of VCC~2034 is not evident. AGC~226178 itself is a peculiar `almost dark' system comparable to SECCO~1, and immediately adjacent to it, but at a completely different distance, is a highly unusual dwarf spheroidal (NGVS~3543). We thus emphasize caution when interpreting these systems, and the importance of exhaustive follow-up observations to eliminate possible interpretations.

\begin{acknowledgments}
The authors thank the anonymous referee for their constructive comments. We also thank Lodovico Coccato and Sungsoon Lim for helpful discussions, and Kyle Artkop for assistance in identifying blue candidates in Virgo.
This work is based on observations made with the NASA/ESA Hubble Space Telescope, obtained at the Space Telescope Science Institute, which is operated by the Association of Universities for Research in Astronomy, Inc., under NASA contract NAS5-26555.  These observations are associated with program \# HST-GO-15183.  Support for program \# HST-GO-15183 was provided by NASA through a grant from the Space Telescope Science Institute, which is operated by the Association of Universities for Research in Astronomy, Inc., under NASA contract NAS5-26555.
It is also based on observations collected at the European Organisation for Astronomical Research in the Southern Hemisphere under ESO programme 0101.B-0376A.
This work used archival data from the Karl G. Jansky Very Large Array. The National Radio Astronomy Observatory is a facility of the National Science Foundation operated under cooperative agreement by Associated Universities, Inc. The data were observed as part of program 13A-028 (PI: J.~Cannon).
The work used images from the Dark Energy Camera Legacy Survey (DECaLS; Proposal ID 2014B-0404; PIs: David Schlegel and Arjun Dey). Full acknowledgment at \url{https://www.legacysurvey.org/acknowledgment/}.
Funding for the SDSS and SDSS-II has been provided by the Alfred P. Sloan Foundation, the Participating Institutions, the National Science Foundation, the U.S. Department of Energy, the National Aeronautics and Space Administration, the Japanese Monbukagakusho, the Max Planck Society, and the Higher Education Funding Council for England. The SDSS Web Site is \url{http://www.sdss.org/}.

DJS acknowledges support from NSF grants AST-1821967 and 1813708.
MB acknowledges the financial support to this research from the INAF Main Stream Grant 1.05.01.86.28 assigned to the program {\em The Smallest Scale of the Hierarchy (SSH)}
KS acknowledges support from the Natural Sciences and Engineering Research Council of Canada (NSERC).
BMP is supported by an NSF Astronomy and Astrophysics Postdoctoral Fellowship under award AST-2001663.
EAKA is supported by the WISE research programme, which is financed by the Dutch Research Council (NWO).
GB acknowledges financial support through the grant (AEI/FEDER, UE) AYA2017-89076-P, as well as by the Ministerio de Ciencia, Innovación y Universidades, through the State Budget and by the Consejería de Economía, Industria, Comercio y Conocimiento of the Canary Islands Autonomous Community, through the Regional Budget.
JS acknowledges support from the Packard Foundation.
MPH acknowledges support from NSF/AST-1714828 and grants from the Brinson Foundation.
JMC, JF, and JLI are supported by NST/AST 2009894.
R.R.M. gratefully acknowledges support from the project ANID PIA/BASAL FB210003. 
Research by DC is supported by NSF grant AST-1814208.

\facilities{Arecibo, Blanco, GALEX, HST (ACS), VLA, VLT:Yepun (MUSE)}
\software{\href{http://americano.dolphinsim.com/dolphot/}{\texttt{DOLPHOT}} \citep{Dolphin2000}, \href{https://casa.nrao.edu/}{\texttt{CASA}} \citep{CASA}, \href{https://www.astropy.org/index.html}{\texttt{astropy}} \citep{astropy2013,astropy2018}, \href{https://aplpy.github.io/}{\texttt{APLpy}} \citep{aplpy2012,aplpy2019}, \href{https://photutils.readthedocs.io/en/stable/}{\texttt{Photutils}} \citep{photutils}, \href{https://reproject.readthedocs.io/en/stable/}{\texttt{reproject}} \citep{reproject}, \href{https://sites.google.com/cfa.harvard.edu/saoimageds9}{\texttt{DS9}} \citep{DS9}, \href{https://dustmaps.readthedocs.io/en/latest/}{\texttt{dustmaps}} \citep{Green2018}, \href{https://astroalign.readthedocs.io/en/latest/}{\texttt{Astroalign}} \citep{astroalign}, \href{https://www.astromatic.net/software/sextractor/}{\texttt{Sextractor}} \citep{Bertin+1996}, \href{https://aladin.u-strasbg.fr/}{\texttt{Aladin}} \citep{Aladin2000,Aladin2014}}

\end{acknowledgments}

\bibliography{refs}{}
\bibliographystyle{aasjournal}



\end{document}